\documentclass[pdflatex,sn-mathphys-ay,iicol]{sn-jnl}

\usepackage{graphicx}%
\usepackage{multirow}%
\usepackage{amsmath,amssymb,amsfonts}%
\usepackage{amsthm}%
\usepackage{mathrsfs}%
\usepackage[title]{appendix}%
\usepackage{xcolor}%
\usepackage{textcomp}%
\usepackage{manyfoot}%
\usepackage{booktabs}%
\usepackage{algorithm}%
\usepackage{algorithmicx}%
\usepackage{algpseudocode}%
\usepackage{listings}%

\theoremstyle{thmstyleone}%
\theoremstyle{thmstyletwo}%

\theoremstyle{thmstylethree}%

\begin{document}

\title[Article Title]{Efficiency of viscous angular momentum transport in dissipating Be binaries}


\author*[1]{\fnm{Peter} \sur{Quigley}}\email{pquigley@uwo.ca}

\author[1]{\fnm{Carol E.} \sur{Jones}}\email{cejones@uwo.ca}

\author[2]{\fnm{Kenneth} \sur{Gayley}}\email{kenneth-gayley@uiowa.edu}

\author[3,4]{\fnm{Anahi} \sur{Granada}}\email{agranada@unrn.edu.ar}

\author[5]{\fnm{Stan} \sur{Owocki}}\email{owocki@udel.edu}

\author[1]{\fnm{Rina} \sur{Rast}}\email{krast@uwo.ca}

\author[1]{\fnm{Mark} \sur{Suffak}}\email{msuffak@uwo.ca}

\author[6]{\fnm{Atsuo T.} \sur{Okazaki}}\email{okazaki@hgu.jp}

\author[7]{\fnm{Asif} \sur{ud-Doula}}\email{asif@psu.edu}

\author[8]{\fnm{Ji\v r\'\i} \sur{Krti\v{c}ka}}\email{krticka@physics.muni.cz}

\author[9]{\fnm{Alex C.} \sur{Carciofi}}\email{carciofi@astro.iag.usp.br}

\author[10]{\fnm{Jeremy J.}\sur{Drake}}\email{jeremy.1.drake@lmco.com}

\affil*[1]{\orgdiv{Department of Physics and Astronomy}, \orgname{University of Western Ontario}, \orgaddress{\street{1151 Richmond St.}, \city{London}, \postcode{N6A 3K7}, \state{Ontario}, \country{Canada}}}

\affil[2]{\orgdiv{Department of Physics and Astronomy}, \orgname{University of Iowa}, \orgaddress{\street{203 Van Allen Hall}, \city{Iowa City}, \postcode{ 52242}, \state{Iowa}, \country{USA}}}

\affil[3]{\orgdiv{Centro Interdisciplinario de Telecomunicaciones, Electrónica, Computación y Ciencia Aplicada - Sede Andina}, \orgname{Universidad Nacional de R\'io Negro}, \orgaddress{Anasagasti 1463}, \city{San Carlos de Bariloche}, \postcode{8400}, \state{R\'io Negro}, \country {Argentina}}

\affil[4]{\orgname{Consejo Nacional de Investigaciones
Científicas y Técnicas}, \orgaddress{ Godoy Cruz 2290}, \city{Ciudad Aut\'onoma de Buenos Aires}, \postcode{C1425FQB}, \country{Argentina}}

\affil[5]{\orgdiv{Bartol Research Institute}, \orgname{University of Delaware}, \orgaddress{\street{210 S College Ave}, \city{Newark}, \state{DE} \postcode{19716}, \country{USA}}}

\affil[6]{\orgdiv{Center for Development Policy Studies}, \orgname{Hokkai-Gakuen University}, \orgaddress{\street{4-1-40 Asahimachi, Toyohira-ku}, \city{Sapporo}, \postcode{062-8605}, \country{Japan}}}

\affil[7]{\orgdiv{Department of Physics}, \orgname{Penn State Scranton}, \orgaddress{\street{120 Ridge View Drive}, \city{Dunmore}, \postcode{18512}, \state{PA}, \country{USA}}}

\affil[8]{\orgdiv{Department of Theoretical Physics and Astrophysics, Faculty of Science}, \orgname{Masaryk University}, \orgaddress{\street{Kotl\'a\v rsk\'a 2}, \city{Brno}, \postcode{611 37}, \country{Czech Republic}}}

\affil[9]{\orgdiv{Instituto de Astronomia, Geof\'ica e Ci\^encias Atmosf\'ericas}, \orgname{Universidade de S\~ao Paulo}, \orgaddress{\city{ S\~ao Paulo}, \postcode{94304}, \state{SP}, \country{Brazil}}} 

\affil[10]{\orgdiv{Advanced Technology Center}, \orgname{Lockheed Martin}, \orgaddress{\street{3251 Hanover St}, \city{Palo Alto}, \postcode{94304}, \state{CA}, \country{USA}}}

\abstract{Angular momentum transport is a fundamental process shaping the structure, evolution, and lifespans of stars and disks across a wide range of astrophysical systems. Be stars offer a valuable environment for studying viscous transport of angular momentum in massive stars, thanks to their rapid rotation, observable decretion disks, and likely absence of strong magnetic fields. This study analyzes angular momentum loss in 40 Be binary simulations spanning a range of orbital separations and companion masses, using a smoothed-particle hydrodynamics (SPH) code. A novel framework is introduced to define the outer disk edge based on the behavior of the azimuthal velocity, streamlining the analysis of angular momentum transport within the system. Applying this framework reveals that systems with smaller truncation radii tend to reaccrete a larger fraction of their angular momentum during dissipation, thereby inhibiting the stars ability to regulate its surface rotation. Modification of this rate may alter the star's mass-injection duty cycle or long-term evolutionary track. Finally, a subset of the simulations were post-processed using the Monte Carlo radiative transfer code \texttt{HDUST}, generating synthetic observables including H$\alpha$ line profiles, V-band polarization, and UV polarization. Suggestions for observational verification of the dynamical results are demonstrated using the simulated data.}


\maketitle

\section{Introduction} \label{sec:Introduction}

The conservation of angular momentum has long been recognized as a fundamental factor in the formation and evolution of circumstellar disks. Whether originating from a collapsing molecular cloud or through mass transfer in binary systems, the rotational barrier that enables disk formation also presents a challenge to its continued evolution. For the disk to collapse further or accrete onto the central object, angular momentum must be efficiently transported outward. Observations across a range of astrophysical systems \textemdash  including young stellar objects, active galactic nuclei, and mass-transferring binaries \textemdash demonstrate that such angular momentum loss does occur, enabling sustained accretion. This process is not only possible but essential to the long-term development of these systems.

Classical B-emission (Be) stars are a unique case in this regard. They are defined by the presence of prominent Balmer emission lines and broadened photospheric lines in their spectrum, indicative of a dense, ionized disk surrounding the star \citep{riv13}. These disks differ from those around most other massive stars in two key ways.  The first is that a Be disk is fed not from its outer edge, as in mass transferring binaries and other accretion disks, but from its inner edge near the equator of the central star \citep{mai03}. In Be stars, the material at the stellar equator has been measured to rotate at a high fraction of the critical rate where it would be in orbit about the star \citep{zor16}. By some mechanism, possibly non-radial pulsations \citep{and86}, this material is ejected into the disk, thereby supplying the disk with mass and angular momentum. The second difference is the lack of any detectable magnetic fields originating from the star, observationally limited to stellar surface fields below 100 G \citep{nei12, wad16}. Stricter still, \cite{ud-d18} found through numerical simulations that magnetic fields of even 10 G can significantly disrupt the characteristic Keplerian disk structure of Be stars. In the absence of large-scale magnetic fields, the growth of Be star decretion disks is  understood to be driven primarily by viscous processes.

The injection of mass and angular momentum feeds the outward diffusion of material from the stellar surface, analogous to an inverse accretion process, often called decretion \citep{oka02}.  In their seminal paper, \cite{sha73} proposed a framework for accretion in black hole systems in which angular momentum transport is dictated by turbulent viscosity. In this formulation, the kinematic viscosity, $\nu$, is defined to be proportional to both the local sound speed, $c_s$, and the vertical scale height, $H$. A dimensionless scale factor, $\alpha$, then generalizes the efficiency of angular momentum transport in the system as

\begin{equation}
    \nu = \alpha c_s H.
\label{eq:viscosity}
\end{equation}

\noindent Later, \cite{pri81} applied this form of viscosity to accretion disks with great success. \cite{lee91} extrapolated this process to decretion disks to explain the growth of Be disks, forming the foundation of the viscous decretion disk (VDD) model: the most successful model of Be disk physics to date. In the three decades since its inception, the VDD model has been extensively validated both numerically and observationally \citep{oka91, bjo05, vie15, gho18, ras25}.

While many studies focus only on steady-state disks, many Be stars are observed to have non-periodic cycles of both active mass injection (decretion) and quiescence (re-accretion and dissipation). During quiescence, mass injection is either entirely stopped or greatly diminished \citep{gho18}. These phases can last from weeks to years, during which the star's brightness fluctuates due to changes in the surrounding circumstellar material. Given enough time for buildup, these systems can reach a steady state in which the disk observables become unaffected by further accumulation.

Although the direction of mass flow is clear during both decretion and accretion phases, angular momentum does not always follow the same path as the material. During decretion, angular momentum must be transported outward along with the mass to allow the disk to grow. However, once the system enters a quiescent phase, the disk material begins to fall back toward the star. For this infall to occur, angular momentum must continue to be transported outward, away from the star \citep{rim18}. This ongoing outward transport results in a net loss of angular momentum from the stellar surface, providing a mechanism by which the star can regulate its rotational criticality in order to address the evolutionary pressures from core contraction and envelope expansion. 

In isolated Be stars, angular momentum can be transported outward by viscous forces over hundreds of stellar radii before the gas gradually decouples from the disk. However, this extended transport is uncommon among Be stars, as many are members of binary systems with companions ranging from compact objects \citep{bro19, fra19} to hot subdwarfs \citep{wan17, wan18}. In such systems, the gravitational influence of the secondary tidally truncates the disk, establishing a well-defined outer boundary beyond which angular momentum can no longer be efficiently redistributed back to the star.

The unique environment of Be disks provides a valuable opportunity to study angular momentum transport inside the star by observing processes outside the star. Stellar rotation is known to play a fundamental role in shaping a star's internal structure \citep{mey00}, chemical mixing \citep{eks12, geo13,gra13}, and evolutionary path \citep{heg00, heg00b}. However, this process is often difficult to study explicitly since only the outer stellar structure is directly visible, while indirect helioseismology techniques require supplemental constraints. Few stellar environments highlight the importance of rotation more clearly than Be stars, which regulate their near-critical surface velocities by ejecting material into their surroundings. By studying how angular momentum is lost through the disk, we may gain valuable insight into the internal mechanisms that govern angular momentum redistribution within the stellar interior.

In the present work, we present the first model of angular momentum loss efficiency from a Be star in a binary system during quiescence. We begin by detailing the models and tools used in this work in Section~\ref{sec:Methods}. We also give details on the initial configurations of the dynamical simulations and implicit assumptions therein. In Section~\ref{sec:TruncRad}, we define a simple criterion for the truncation radius of the Keplerian disk from the azimuthal velocity profile. Using this definition, we then create a scaling law for intermediate-type Be stars relating the truncation radius to the binary mass ratio and orbital separation. In Section \ref{sec:AngMom}, we apply this framework in order to discuss the rate of angular momentum loss from the Be stars in quiescent systems. In Section~\ref{sec:Observables}, we select a subset of the dynamical models discussed in previous sections to simulate the observable signatures of each disk using the radiative transfer code \texttt{HDUST} \citep{car06}. Using this data, we examine the H$\alpha$ line, V-band polarization, and UV polarization to identify indicators that could enable observational verification of the predicted angular momentum loss rates from the primary star. 

\section{Methodology} \label{sec:Methods}

\subsection{Viscous Decretion Disk Model} \label{sec:VDD}

For the past three decades, the VDD model has served as the primary framework for understanding Be star disks \citep{lee91}. It has since been able to successfully reproduce a wide range of observed phenomena, including Be variability \citep{gho18}, violet-to-red (V/R) variation \citep{oka91}, and disk tearing \citep{suf22}. 

A critical factor in the VDD model's success is its description of the unique mass injection mechanism. In this model, the central Be star is assumed to be rotating near the critical limit where the centrifugal force would be balanced against the gravitational force at the equator. By some process, often attributed to non-radial pulsations \citep{mai03}, mass and angular momentum are injected into the disk at Keplerian speeds according to

\begin{equation}
    v_\phi = v_\mathrm{kep} = \sqrt{\frac{GM}{r}},
    \label{eq:vkep}
\end{equation}

\noindent in a thin, symmetric annulus about the equator \citep{oka91, mei07}. It should be noted that other studies have investigated the possibility of asymmetrical injection zones (such as \citealt{lab25}) but conclude that the injected material is quickly circularized in any case.

Within the VDD model framework, gravity, pressure, and kinematic viscosity are the primary factors governing the disk's dynamics. As discussed in the introduction, viscosity is described following the $\alpha$-parameterization of \cite{sha73} according to Equation~\ref{eq:viscosity}. The $\alpha$ parameter is typically thought to range between 0.1 and 1.0 in Be stars \citep{gho17}.

In the steady state limit, the disk is assumed to be isothermal with a vertical structure in hydrostatic equilibrium. It is typically assumed for Be disks that the gas is non-self-gravitating and so the scale height of the disk is determined only by gas pressure and the gravitational force of the central star as

\begin{equation}
    H(r) = H_0 \left(\frac{r}{R_\star}\right)^{3/2},
    \label{eq:ScaleHeight}
\end{equation}

\noindent where $H_0 = c_s R_\star/v_{orb}$, $R_\star$ is the equatorial radius of the primary star, and $v_{orb} = \sqrt{GM_\star/R_\star}$ is the orbital velocity at the stellar surface, where $M_\star$ is the mass of the primary star. This holds under the thin disk approximation where $H\ll r$ (i.e., near the star) and becomes inaccurate at large radii \citep{kur18}. Due to the prior assumption of isothermality, the sound speed therefore  becomes

\begin{equation}
    c_s^2 = c_{s,\mathrm{iso}}^2 = \frac{kT_d}{m_H \mu},
\end{equation}

\noindent where $k$ is the Boltzmann constant, $T_d$ is the disk temperature, $m_H$ is the hydrogen mass, and $\mu$ is the mean molecular weight of the gas. In Be star disks, the disk temperature is well-approximated as 60\% of the primary star’s effective temperature, $T_\mathrm{eff}$, as suggested by \cite{car06}. The mean molecular weight of the gas near the star is typically taken to correspond to that of an ionized gas with solar metallicity ($\sim0.6$) \citep{car06}.

After a sufficiently long mass-injection phase, the volume density of the disk around an isolated Be star (under the assumptions outlined above) can be approximated by the expression

\begin{equation}
    \rho(r,z) = \rho_0 \left(\frac{r}{R_\star}\right)^{-n} \exp\left(\frac{-z^2}{2H^2}\right),
\label{eq:vdensity}
\end{equation}

\noindent where $z$ is the vertical coordinate, $\rho_0$ is the base volume density in the equatorial plane at the stellar surface and $n$ is a radial density exponent. Note that because of the assumptions made regarding mass injection and viscosity, the disk is inherently axisymmetric. By integration, we can find the equivalent surface density equation as

\begin{equation}
    \Sigma(r) = \Sigma_0\left(\frac{r}{R_\star}\right)^{-(n-1.5)} = \Sigma_0\left(\frac{r}{R_\star}\right)^{-m},
\end{equation}

\noindent where $\Sigma_0$ and $m$ are the base surface density and analogous radial density exponent respectively. When applying the assumptions discussed above to the Navier-Stokes equation, an isothermal disk is found to have radial exponent $n=3.5$ (equivalently $m=2.0$). 

In order to prevent dissipation, mass and angular momentum must be continually supplied to the disk to replace material which is re-accreted onto the star or is transported radially outwards by viscosity. This rate is set by

\begin{equation}
    \Sigma_0 = \frac{(GM_\star)^{1/2} \dot{M}(1-\Xi)}{2\pi\alpha c_s^2} \frac{r_\mathrm{disk}^{1/2} - R_\star^{1/2}}{R_\star^2}
    \label{eq:MIR}
\end{equation}

\noindent where $\dot{M}$ is the mass injection rate, $r_\mathrm{disk}$ is the size of the disk, and

\begin{equation}
    \Xi = \frac{r_\mathrm{disk}^{1/2} - r_\mathrm{inj}^{1/2}}{r_\mathrm{disk}^{1/2} - R_\star^{1/2}}
\end{equation}

\noindent is the fraction of injected mass that flows back to the star, with $r_\mathrm{inj}$ being the radius at which material is injected. These equations have been adapted from \cite{bjo05}, where a full derivation for the disk structure is presented starting from the fluid equations, with modifications to reflect the possibility that some fraction of the injected material is quickly re-accreted.

In binary systems, these equations are complicated significantly by the gravitational influence of the secondary. Particularly, recent works have shown that the vertical structure near the companion is modified, requiring a correction to the scale height \citep{rub25}. Additionally, Be star disks have been observed to have shallower distribution of $n \lesssim 3.0$ due to tidal truncation, typically referred to as the mass accumulation effect \citep{oka02, pan16, vie17}.

\subsection{SPH Implementation}

The simulations presented in the paper were done by a thee-dimensional smoothed particle hydrodynamics (3D SPH) code based on a version originally developed by \cite{ben90} and \cite{bat95} and modified for Be disks by \cite{oka02}. The intent of this implementation of the SPH equations was to have accurate, memory-efficient simulations of single and binary Be star systems on a range of dynamical time-scales. Since its inception, many works have utilized this SPH code to study the dynamics of decretion disk systems \citep{pan16,cyr17, ras25, suf25}. 

In this study, we investigate the dynamics of decretion disks in binary star systems, modeling both the Be star and its companion as spherical sink particles. The accretion radius of each star is defined so that any particle within the star's radius is removed from the simulation. All particles removed this way have the time of accretion recorded, as well as their current position, velocity, and mass. While previous studies often define the secondary sink to be its Roche lobe due to the low gas density in that region (for example \citealt{suf22} and \citealt{cyr17}), this approach would introduce significant bias into our simulations. Since we seek to study the loss of angular momentum from the disk, it is essential that particles be permitted to re-enter the disk from the circumsecondary or circumbinary regions. A Roche lobe sink prevents the formation of these structures, thereby limiting the accuracy of our result. To address this, we adopt the physical radius of the companion as its sink boundary, enabling the formation and interaction of extended disk structures. The simulation itself is bounded at $50\, R_\star$, beyond which particles are assumed to never return to the circumprimary disk and are removed from the simulation.

The properties of each particles are assigned dynamically as the simulation is evolved using a second-order Runge–Kutta–Fehlberg integrator. The position and velocity of each particle is updated at each timestep, while smoothing length is adaptively determined based on the proximity of its nearest neighbors. The sole exception is particle mass, which is statically assigned at the time of injection and is uniform across all particles in the simulation.

As previously stated, $\alpha$ is assumed to be constant throughout the disk. However, the code implements an artificial viscosity scheme which is allowed to vary in space and time in order to keep the Shakura \& Sunyaev $\alpha$ fixed. In this work, we incorporate the correction introduced by \citet{rub25}, which accounts for the gravitational influence of both stars in the binary system. A description of the SPH equations and of artificial viscosity can be found in \cite{mon05}.

When mass injection is turned on, particles are spawned into a thin annulus at the equator of the star, at a radius of $r_\mathrm{inj} = 1.04R_\star$. They are launched in circular Keplerian orbits, as defined by Equation~\ref{eq:vkep}, relative to the velocity of the primary star. The number of particles injected at each timestep is determined by the mass loss rate divided by the fixed particle mass, ensuring that all particles in the simulation have uniform mass. 

\subsection{Initial Distribution \& Disk Generation}

At the start of each simulation, a Keplerian disk consisting of 200,000 particles is initialized. The disk is populated using a rejection sampling algorithm to ensure that the resulting density profile follows Equation~\ref{eq:vdensity} with $n=3.5$, extending out to a specified outer boundary. This setup differs from other uses of the SPH code which typically start with no gas particles and build a disk through continual injection over tens or hundreds of periods until a steady state is reached (e.g., \citealt{oka02, suf22, rub25}). Using this approach, a prebuilt disk can achieve a steady state within one to five orbital periods for an $\alpha=1.0$. This saves hundreds of hours of computational time per simulation and allows for a specific density prescription of the inner disk.

For each disk model, the outer radius of the initial distribution was chosen to be 3/4 of the binary's semi-major axis, $a$, to ensure adequate resolution throughout the disk without presupposing a truncation radius. Particles initially located within the Roche lobe of the secondary are rapidly removed by tidal interactions, either accreted by the secondary or ejected into the circumbinary disk. A larger radius was avoided to prevent immediate large-scale accretion onto the secondary, thereby minimizing early resolution loss that would persist throughout the simulation.

\begin{figure}[t]
\centering
\includegraphics[width=\columnwidth]{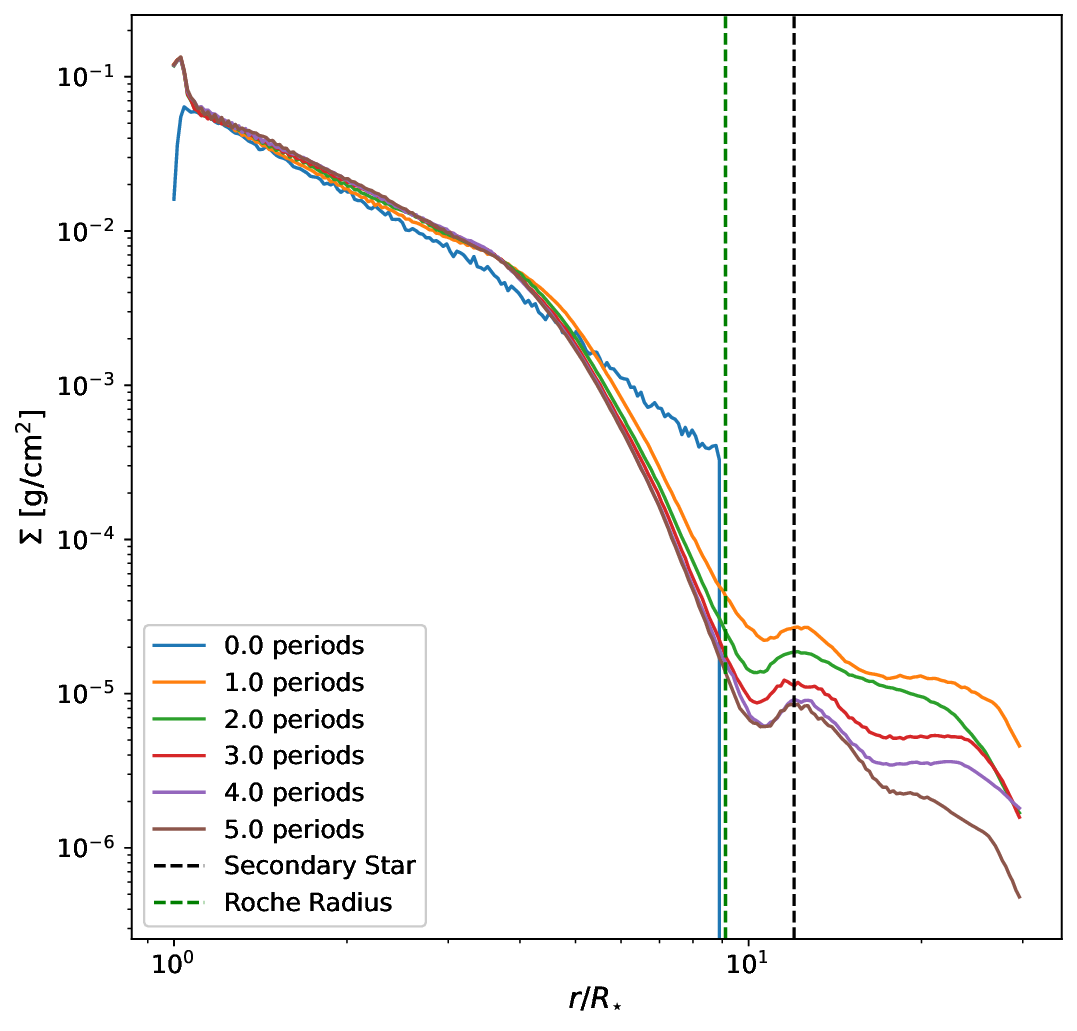}
\caption{Azimuthally-averaged surface density profile from a representative mass-injecting simulation, shown evolving over time until convergence. The initial disk (blue line) is populated only up to 3/4 the semi-major axis. The densest region converges within one period, while the area between this region and the secondary's orbiting radius converges within five orbital periods.}\label{fig:SDconverge}
\end{figure}

While the derivation of Equation~\ref{eq:vdensity} assumes an isolated Be star in steady state, this solution nonetheless provides a suitable initial condition for simulations of a Be star in a binary system from which the disk can evolve and relax toward a steady-state configuration. Following the initial setup, we inject mass at a rate given by Equation~\ref{eq:MIR}, using $r_\mathrm{disk}$ as the outer boundary of the initial disk. The simulation is evolved until the azimuthally-averaged surface density from the surface of the primary star to the Roche lobe of the secondary varies by less than 5\% over a single orbital period. Each simulation in this paper will assume an $\alpha=1$, and with this choice convergence is observed to take at most five periods. However, the densest region becomes well-converged within one orbital period. We chose this value of $\alpha$ in order to examine the fastest-evolving disk (since viscous timescales scale inversely proportional to $\alpha$) to further save on computational time. Lower values are expected to evolve the disk similarly, though at a slower rate and with stronger truncation. A representative surface density profile of a mass-injecting disk can be seen in Figure~\ref{fig:SDconverge}.

The particle mass is determined by the choice of $\rho_0$ in the initial disk setup. By integration of the volume density equation for radial bounds $R_\star~<r~<r_\mathrm{disk}$, we find

\begin{align}
    M_\mathrm{disk} &= \int_{R_\star}^{r_\mathrm{disk}} \int_{-\infty}^{\infty} \int_{0}^{2\pi} \rho_0 \left(\frac{r}{R_\star}\right)^{-3.5} e^{-z^2/2H^2} r\ d\phi\ dz\ dr ,\\
    &= (2\pi)^{3/2}\rho_0  R_\star^2 H_0 \ln{\left(\frac{(3/4)a}{R_\star}\right)},
\end{align}

\noindent All particles, whether in the initial distribution or injected after, are then assigned a particle mass equal to $M_\mathrm{disk}/200\,000$. 

Following the five orbital periods allocated for convergence, mass injection is set to zero. We define the moment between phases as $t=0$. This allows for continuity between a dissipating disk and its long-building precursor, without requiring a detailed and computationally expensive construction phase.

\section{Truncation Radius}
\label{sec:TruncRad}

Several criteria have been proposed to define the truncation radius of the disk. \cite{pan16} modeled a number of simple circular, coplanar binaries and examined their surface density profiles. They found that the profiles were well-fit to a broken powerlaw, and proposed that the truncation radius be defined where the break occurs. This point can be seen in Figure~\ref{fig:SDconverge}, where the slope changes near $4R_\star$. Recently, \cite{rub25} have submitted a new taxonomy for Be disks. Instead of a single truncation radius, they define an inner disk which is nearly unaffected by the tidal influence of the secondary. As radius increases, the perturbation increases until the azimuthal density variation exceeds 5~\% at which point we enter the ``spiral dominated" disk. This region is elongated along the Roche potentials of the binary system in their simulations, giving them their outer disk boundary.

In this section, we will describe an alternative criterion for the truncation radius, which is chosen specifically to benefit our analysis of angular momentum transport. In contrast to \cite{pan16} and \cite{rub25}, who defined their truncation radii based on characteristics of the surface density, we adopt a criterion based on the azimuthally-averaged azimuthal velocity. We argue that this alternative framework is a convenient and effective tool for measuring and analyzing the angular momentum content of a disk.

\subsection{Truncation Criterion}

Even for a Be star in a binary system, the innermost region of the circumprimary disk is expected to follow a Keplerian rotation curve. This behavior arises because, close to the star, the gravitational influence of the secondary is relatively weak compared to that of the primary and is insufficient to counteract the circularizing effects of viscous forces \citep{rub25}. Less obvious, however, is the location within the disk where tidal forces from the secondary begin to dominate and significantly alter the orbits of the disk material.

\begin{figure}[t]
\centering
\includegraphics[width=\columnwidth]{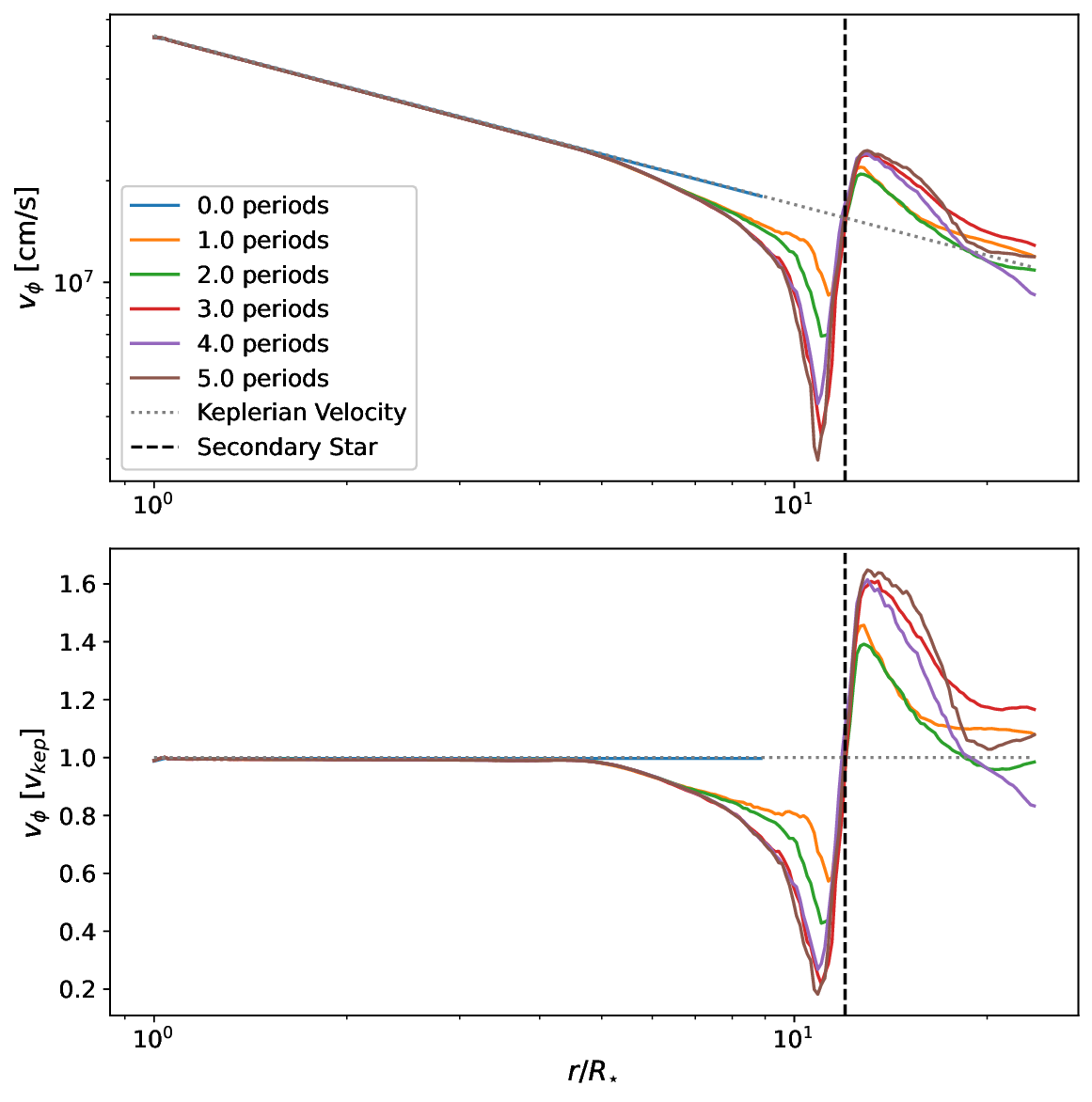}
\caption{Azimuthally-averaged azimuthal velocity of a sample system during the five mass-injecting orbital periods. The top panel shows the velocity in \texttt{cgs} units, while the bottom panel presents the velocity normalized to the local circular speed, $v_\mathrm{kep}$. The orbiting radius of the companion is shown as a black vertical line at $12R_\star$.}\label{fig:VphiEx}
\end{figure}

Figure~\ref{fig:VphiEx} presents the azimuthally-averaged azimuthal velocity, $v_\phi$, of a sample system during the build-up phase. To capture both the standard view and a comparison to the theoretical model, $v_\phi$ is plotted both in CGS units and as a ratio of the circular Keplerian velocity as described in Equation~\ref{eq:vkep}. 

The top plot replicates the $v_\phi$ profile of \cite{rub25}. In that study, the disks were built from initially empty systems over many mass-injection periods, lending credibility to the alternative steady-state initialization method employed in hte present simulations.

The lower panel, which presents the velocity normalized to $v_\mathrm{kep}$, can be divided into two regions. From the stellar surface, a circular disk extends outwards, terminating in a sharp decrease at some point inside the secondary's orbiting radius. In this inner region, the primary's gravitational influence dominates, and viscous forces efficiently redistribute any angular momentum imparted by the companion. Beyond the downturn, however, tidal effects from the companion overwhelm the viscous response, causing the particles' orbits to elongate and deviate from $v_\mathrm{kep}$. The boundary between regions naturally gives rise to the truncation radius condition,

\begin{equation}
    r_\mathrm{trunc}: v_\phi(r) / v_\mathrm{kep}(r) = 98\%.
    \label{eq:rtrunc}
\end{equation}

\noindent The 98\% threshold has been chosen to allow only minor deviations from a circular orbit, without truncating the disk so severely that significant quantities of mass and angular momentum are excluded. However, there is some tolerance for adjustment without major differences. For example, at the end of convergence in the sample simulation shown in Figure~\ref{fig:VphiEx}, increasing the threshold to 99\% shifts $r_\mathrm{trunc}$ from 5.1 $R_\star$ to 3.3 $R_\star$ and significantly decreases the amount of mass and angular momentum included in the disk by 23\% and 30\%, respectively. Relaxing the threshold to 95\%, in contrast, moves $r_\mathrm{trunc}$ outwards to 5.7 $R_\star$, but results in only a modest increase in the mass and angular momentum (3\% and 4\%, respectively). 

The purpose of defining the truncation radius, which we will refer to as the $v_\phi$ turnoff to distinguish it from other criteria, is twofold. The first advantage is that it allows us to directly relate the angular momentum to the density of the disk as,

\begin{align}
    J_z &= 2\pi\int_\mathrm{R_\star}^{R_\mathrm{out}} \Sigma (r) v_\phi(r) r^2 dr, \\
    &\approx 2\pi\int_\mathrm{R_\star}^{r_\mathrm{trunc}} \Sigma (r) v_\mathrm{kep}(r) r^2 dr.
    \label{eq:InnerDiskAngMom}
\end{align}

\noindent However, no restrictions are placed on disk morphology. This allows for a single condition to characterize building, steady-state, and dissipating disks without modification. In fact, as we will discuss, this is a property not of the disk itself, but purely of the characteristics of the binary. 

The other advantage is that measurement of the $v_\phi$ turnoff can be done using different observables than have been historically used. Disk truncation has traditionally been inferred from the turndown in the spectral energy distribution (SED) \citep{kle19}, although such data may not always be available for a given star. However, as we will discuss in Section~\ref{sec:Observables}, the truncation radius may also be estimated under specific circumstances using the H$\alpha$ profile, a common, low-cost observable, or using the V-/UV-polarization, the latter of which is the focus of the proposed Polstar mission \citep{jon22}.

In summary, the $v_\phi$ turnoff radius acts to simplify the angular momentum integral in two ways. Firstly, it replaces what could be a complicated velocity term by a simple prescriptive $v_\mathrm{kep}$. Secondly, it provides a natural upper bound to the integral, which may otherwise be contentious in a dissipating system where the powerlaw breakpoint is not always clearly identifiable and excludes some angular momentum at larger radii.

\subsection{Simulation Details}

In each simulation, we adopt a B5e-type star as the primary, with stellar parameters consistent with the representative values reported by \cite{suf23}, which are themselves based on tabulations by \cite{cox00}. Specifically, we set the stellar mass to $M_\star~=~5.9 M_\odot$, radius to $R_\star~=~3.9 R_\odot$, effective temperature $T_p = 15000\textrm{ K}$, and disk base density $\rho_0~=~5~\times~10^{-12}~\textrm{ g/cm}^3$. An intermediate-type B star was chosen so that stellar winds, which have not been included in our models, do not play a significant role in shaping the disk structure or in angular momentum transport \citep{kee16, gra13}. 

The mass of the secondary star is varied across simulations, while its radius is held fixed at $R_s = 1.0R_\odot$. This value is physically representative of typical subdwarf companions, which are frequently observed to accompany Be stars \citep{wan17, wan18}. From a modeling perspective, this fixed radius represents a balance between two competing requirements: it is sufficiently large to enable efficient accretion of material overflowing the primary's Roche lobe, yet small enough to avoid artificially obstructing gas that might otherwise enter a circumbinary orbit. The secondary mass is sampled from the set: 0.2, 0.33, 0.5, 0.75, 1.0, 1.5, 2.0, and 3.0 $M_\odot$, encompasing a wide range of companion types, from white dwarfs to late-type B stars. This range captures a physically meaningful diversity of systems while remaining computationally tractable, particularly given the limited resolution in the circumsecondary region.

The stars are set in circular orbits with the semi-major axis varied across simulations. Expressed in units of the primary star's radius, the orbital separation of the two stars is set to 8, 10, 12, 14, or 16 $R_\star$. For orbital separations less than $8 R_\star$, the disk was found to truncate very close to the primary star, leading to dissipation within just a few days. As a result, these models are excluded from the present analysis due to their limited dynamical evolution. For orbital separations greater than $16 R_\star$, the long orbital period causes the secondary-disk interaction time to shorten, weakening the overall effect of tidal torques by the secondary. This selection therefore represents a sizeable parameters space where the tidal torques are not strong enough to rapidly destroy the disk but not weak enough to have no effect.

Once mass injection is stopped, each dissipating disk is allowed to evolve for 20 orbital periods, after which the simulation is terminated. This timescale is sufficient to capture the key features of disk dissipation while avoiding the loss of resolution that occurs as particles are gradually removed from the simulation.

\subsection{Truncation Evolution}

Although all particles are initially in circular orbits when the mass injecting simulations start, the disk quickly evolves into a distinct circular inner region and non-Keplerian, low-density outer region. For the example simulation shown in Figure~\ref{fig:VphiEx}, the division between the region develops within one orbital period. Even at this early time, the $v_\phi$ turnoff can be clearly identified and remains constant for the entire settling time after.

\begin{figure*}[t]
\centering
\includegraphics[width=\textwidth]{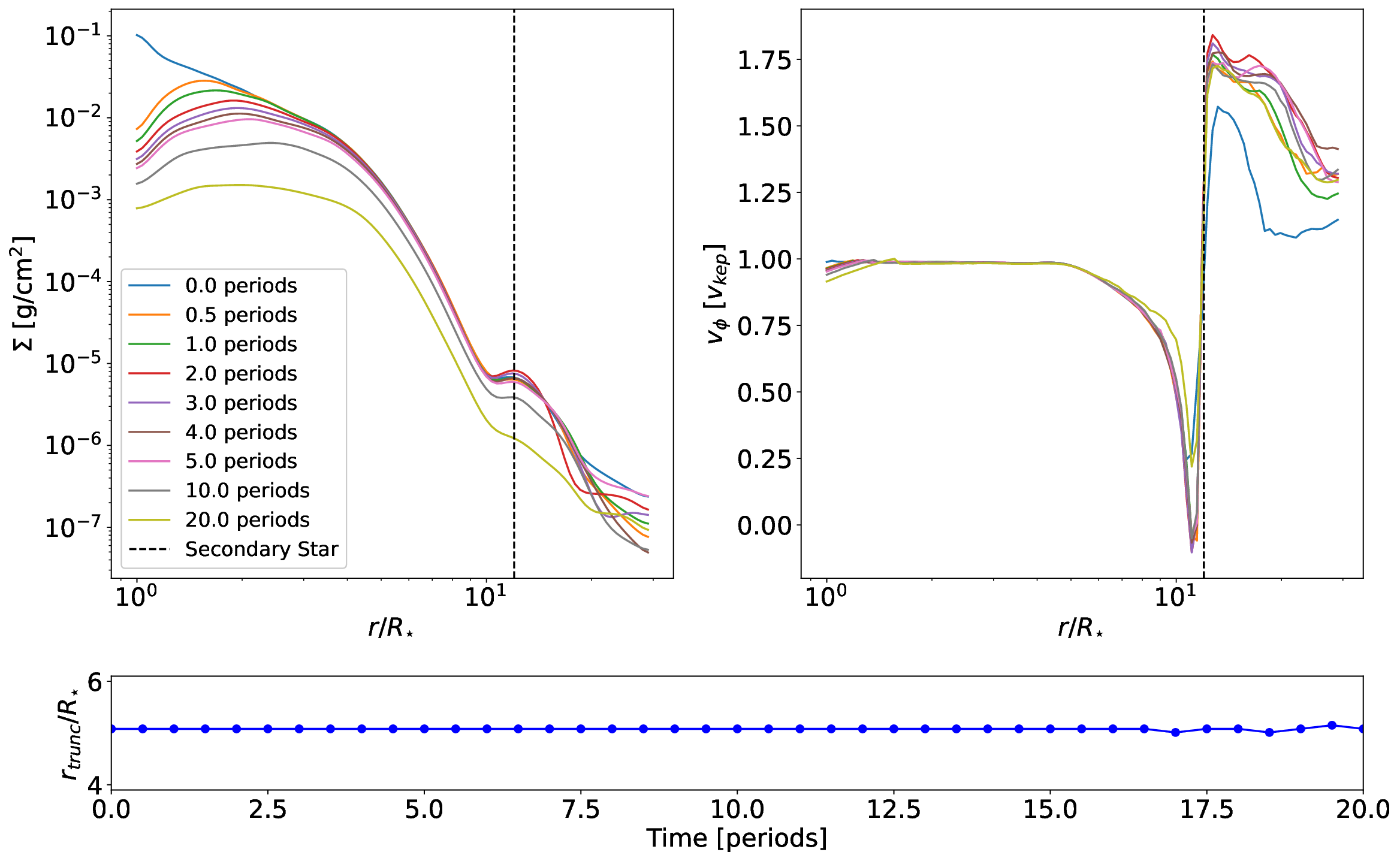}
\caption{Evolution of the azimuthally-averaged surface density (top left) and azimuthal velocity (top right) for a representative simulation of a dissipating disk. The bottom panel tracks the evolution of the $v_\phi$ turnoff radius over the same period. This simulation corresponds to an orbital separation of $a = 12R_\star$ and a secondary mass of $M_s = 1.0M_\odot$.}\label{fig:TurnoffEvol}
\end{figure*}

Once dissipation begins, the morphology of the surface density profile significantly changes from the broken power law described in \cite{pan16}. The disk begins to empty from re-accretion of particles onto the primary and from outflowing particles through the outer edge. Particles exiting the disk can either be accreted by the binary companion or become a part of the circumsecondary or circumbinary disks. This results in a ``humped" surface density distribution, significantly different in shape and magnitude to its initial configuration, as shown in the top left panel of Figure~\ref{fig:TurnoffEvol}.

In comparison to the surface density, the $v_\phi$ distribution changes very little through dissipation. A minor deviation is observed at the inner edge, which spins down as the disk dissipates. The size of this change is as much as 15\% after 20 periods in the sample simulation shown in the top right panel of Figure~\ref{fig:TurnoffEvol} and extends from the stellar surface to 1.75 $R_\star$. It can be inferred that with no angular momentum being injected into the inner disk, material near the stellar surface donates its angular momentum to larger radii but does not receive from smaller radii in kind. As a result, these particles do not have sufficient angular momentum to stay in orbit, and with a sub-circular azimuthal speed begin to spiral inwards towards the star. Even so, when compared to a two-magnitude change in the surface density plot at the same radii, it can be concluded that redistribution of angular momentum through viscous action occurs much more rapidly than does the redistribution of mass.

Throughout the dissipation phase, the distribution at and beyond the $v_\phi$ turnoff remains largely unchanged, even as surface densities decrease by an order of magnitude. Only minor fluctuations are observed at later times, particularly at large periods. Notably, the bottom panel of Figure~\ref{fig:TurnoffEvol} shows that the measured $v_\phi$ turnoff exhibits only slight variation after 17 orbital periods. These small changes may reflect reduced resolution at these later stages, rather than genuine dynamical evolution. This stability allows us to treat the $v_\phi$ turnoff as effectively constant over time, enabling the definition of a static truncation point. We can then use this point to analyze the angular momentum contained `inside' the disk without needing to consider the time-dependant behaviour of its outer edge. Furthermore, its apparent independence from significant changes in local density and radial structure suggests that the turnoff location is primarily determined by the binary configuration and the viscous strength of the disk.

\subsection{Impact of Orbital Parameters}\label{sec:ScaleLaw}

\begin{figure*}[t]
\centering
\includegraphics[width=\textwidth]{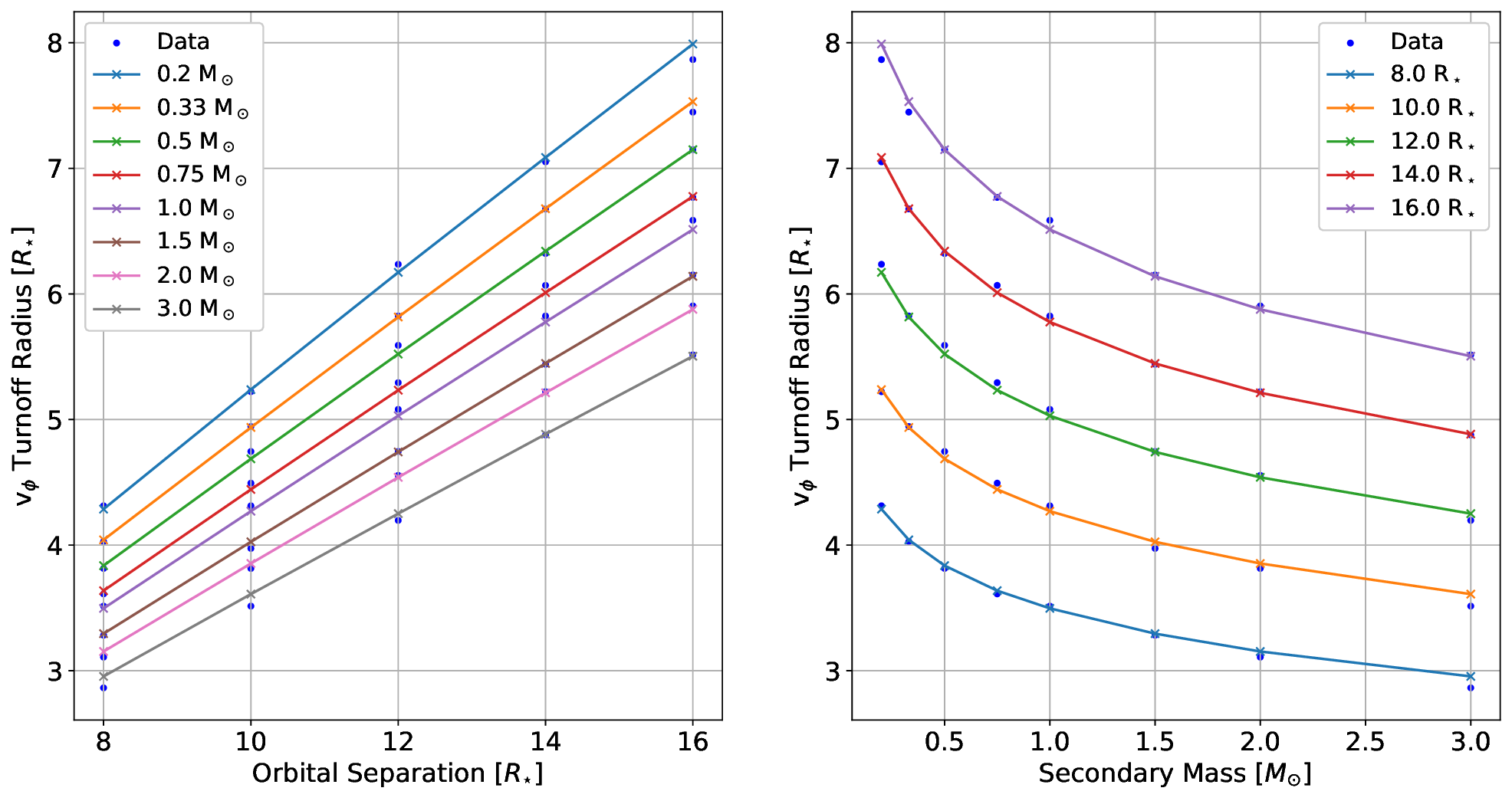}
\caption{Azimuthal velocity $v_\phi$ turnoff for the simulation grid, each measured at $t=10P_\mathrm{orb}$ plotted as blue circles, and the best-fit scaling law. The largest residuals occur for the simulations with large orbital separations or largest secondary masses.}\label{fig:ScaleLaw}
\end{figure*}

From our models, we are able to create an empirical scaling law as a function of secondary mass, $M_s$, orbital separation, $a$, and a proportionality constant, $\beta$, given as,

\begin{equation}
    \frac{r_\mathrm{trunc}}{R_\star} = \left(\beta - \gamma \log_{10}\left(\frac{M_s}{M_\star}\right) \right) \left(\frac{a}{R_\star}\right)^\eta ,
    \label{eq:scalelaw}
\end{equation}

\noindent where $\gamma$ and $\eta$ are free parameters. By applying a non-linear least squares algorithm, we found best-fit values of $\beta=0.391 \pm 0.008$,  $\gamma=0.177\pm0.004$, and $\eta=0.908\pm0.007$. The errors are assumed from the covariance matrix. Each of $r_\mathrm{trunc}$, $M_s$, and $a$ are scaled to properties of the primary star. The raw data and overplotted fits are shown in Figure~\ref{fig:ScaleLaw}.

The form of Equation~\ref{eq:scalelaw} was chosen intentionally so that it can be conveniently compared to the Roche lobe of the primary. For $q = M_s/M_\star \lesssim 0.5$, \cite{pac71} presents a Roche lobe approximation of the form

\begin{equation}
    \frac{R_L}{a} \approx 0.38 - 0.2 \log_{10}(q).
    \label{eq:RL_alternative}
\end{equation}

\noindent which is directly analogous to Equation~\ref{eq:scalelaw}.

One may expect that the $v_\phi$ turnoff should scale linearly with orbital separation, given that the gravitational influence of both stars would fall off equally as $r^{-2}$. This reasoning indeed holds for the Roche lobe. However, our fitted power-law index $\eta < 1.0$ indicates that the truncation radius grows sub-linearly with orbital separation. Consequently, wider binaries do not host proportionally larger disks, implying that other dynamical effects, such as tidal torques or viscosity, play a significant role in limiting disk extent.

The relationship for the companion is slightly more complex. As the companion becomes more massive, the tidal torque it exerts on a given annulus of disk material strengthens, removing angular momentum from the disk and torquing material inwards. When the orbit of the torqued fluid element cannot be efficiently recircularized by viscous processes within a single orbital period, the disk becomes distorted, and the azimuthally averaged azimuthal velocity deviates from $v_\mathrm{kep}$. 

A comparison between the fit parameters in the mass-dependent term of Equation~\ref{eq:scalelaw} and Equation~\ref{eq:RL_alternative} reveals a marked similarity. $\beta=0.391$, for instance, is only slightly larger than the accepted Roche scaling of 0.38. This slight increase introduces a small systematic offset between the two formulations, favoring the scale law by predicting a marginally larger truncation radius. This effect is partially mitigated by the relatively small value of $\gamma < 0.2$, which governs the sensitivity of the truncation radius to variations in the mass ratio, $q$. The first derivative of the truncation term with respect to $q$, given by $-\gamma\, q^{-1}$, indicates that a smaller $\gamma$ results in a flatter dependence of the truncation radius on $q$. Consequently, the scale law exhibits a more gradual decline in $v_\phi$ turnoff with increasing $q$ compared to the Roche lobe model.

These two opposing effects give rise to distinct regimes. When $q > 0.33$, the gravitational influence of the secondary star truncates the $v_\phi$ turnoff less severely under the scale law than under the Roche lobe approximation. Conversely, for $q < 0.33$, the scale law predicts a more conservative truncation, with the $v_\phi$ turnoff occupying less of the interbinary space than the Roche lobe would suggest. This nuanced behaviour underscores the importance of $\beta$ and $\gamma$ in shaping the truncation profile and highlights that even small deviation in parameters values can lead to meaningful differences in physical interpretation.

The close resemblance between the scale law and the Roche lobe approximation implies that gravitational equipotentials play a central role in shaping the truncation behavior. Since the Roche lobe is defined by a critical gravitational potential surface, the similarity in functional form suggests that the scale law is effectively capturing the underlying potential structure of the binary system. This reinforces the interpretation that the truncation of the $v_\phi$ turnoff is governed not merely by geometric constraints, but by the topology of the gravitational field itself.

It should be noted that the truncation radii shown in Figure~\ref{fig:ScaleLaw} scale similarly to the radii listed in Table 1 of \cite{pap77}, where the tidal torques are in balance with the viscous effects.

\section{Angular Momentum Analysis}
\label{sec:AngMom}

To analyze the flow of angular momentum from a dissipating Be disk, we will use the simulations presented in Section~\ref{sec:TruncRad}. We classify the particles in the simulation as ``in the disk" (if $R_\star \le r \le r_\mathrm{trunc}$) or ``outside of the disk" (if $r > r_\mathrm{trunc}$). We can then define the total quantity of angular momentum inside the disk, $\vec{J}_\mathrm{disk}$, within the static $v_\phi$ turnoff radius using Equation~\ref{eq:InnerDiskAngMom} and the time derivative of that quantity as,

\begin{equation}
    \frac{\partial J_\mathrm{disk,z}}{\partial t} = 2\pi\int_\mathrm{R_\star}^{r_\mathrm{trunc}} \frac{\partial\Sigma (r,t)}{\partial t} v_\mathrm{kep}(r) r^2 dr\,,
    \label{eq:Jdot}
\end{equation}

\noindent where $\Sigma$ is now also a function of time due to dissipation. Since the disk and binary orbit are coplanar, we will only consider the $z$-component of angular momentum as the others are small in comparison. The analysis is done in a non-rotating, co-moving reference frame centered on the primary star, such that all quantities are relative to the Be star. This choice allows us to isolate the angular momentum introduced into the disk by the Be phenomenon from the orbital angular momentum of the binary system.

Once dissipation begins, the particles in the disk can be removed by either accretion onto the primary star, $J_\mathrm{acc}$, or by flowing through the $v_\phi$ turnoff boundary, $J_\mathrm{lost}$. Once particles are transported beyond the outer boundary, they are no longer tracked by this analysis. In doing so, we remove any concerns about resolution in the outer regions. These particles and their associated angular momentum are considered lost to the Be star and contribute to the removal of angular momentum from the stellar surface.

Angular momentum can also be lost to the Be star by gravitational torques from the secondary on the inner disk, which does not inherently change with the mass of the disk. However, since the disk volume was chosen such that the azimuthal velocity is near-Keplerian, the viscous action in this region is able to transport the angular momentum outwards quickly by construction. This allows us to combine the angular momentum loss by tidal torques and that lost by mass outflow into a single quantity, $J_\mathrm{lost}$, and hence further simplify our analysis.

\begin{figure}[t]
\centering
\includegraphics[width=\columnwidth]{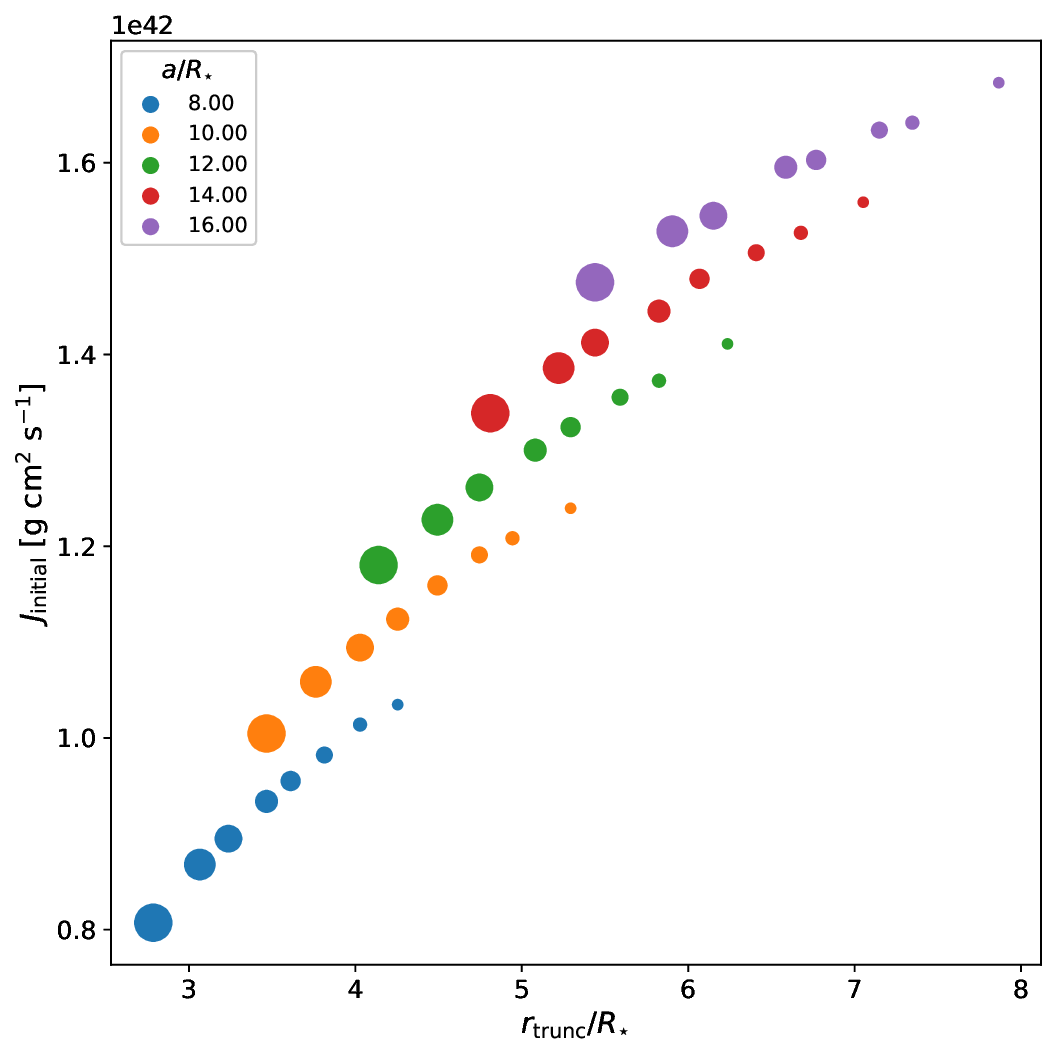}
\caption{Total angular momentum of the disk at $t=0$ as a function of $v_\phi$ turnoff radius. Different orbital separations are differentiated by marker colour. The area of each point is proportional to the mass of the secondary.}\label{fig:InitialJdot}
\end{figure}

Figure~\ref{fig:InitialJdot} shows the total angular momentum in the disk at the instant mass injection is turned off as a function of $v_\phi$ turnoff. As noted in the previous section, more massive companions cause the disk to truncate closer to the primary star and, as such, the $v_\phi$ turnoff increases as the secondary mass decreases for a fixed orbital separation. Generally, larger disks have more angular momentum as more mass is contained within the boundary. However, $J_\mathrm{initial}$ does not scale linearly with $r_\mathrm{trunc}$, and instead becomes less sensitive to $r_\mathrm{trunc}$ as the disk size increases relative to the orbital separation. This effect occurs because the density at large radii becomes very low and thereby contributes very little mass and angular momentum to the disk overall.

Where two disks would have a similar $r_\mathrm{trunc}$, the system with a larger orbital separation always has a larger quantity of angular momentum. This is because larger orbital systems have more massive disks which can store larger quantities of angular momentum. The difference in disk masses occurs because of different surface density break points. As described in \cite{pan16}, the turndown in the surface density profile (see, for example, $4R_\star$ in  Figure~\ref{fig:SDconverge}) occurs at larger radii for smaller or wider binaries. Therefore, the high-density region extends further for such systems and the outer regions of the disk near the $v_\phi$ turnoff contribute more mass.

\begin{figure*}[t]
\centering
\includegraphics[width=0.8\textwidth]{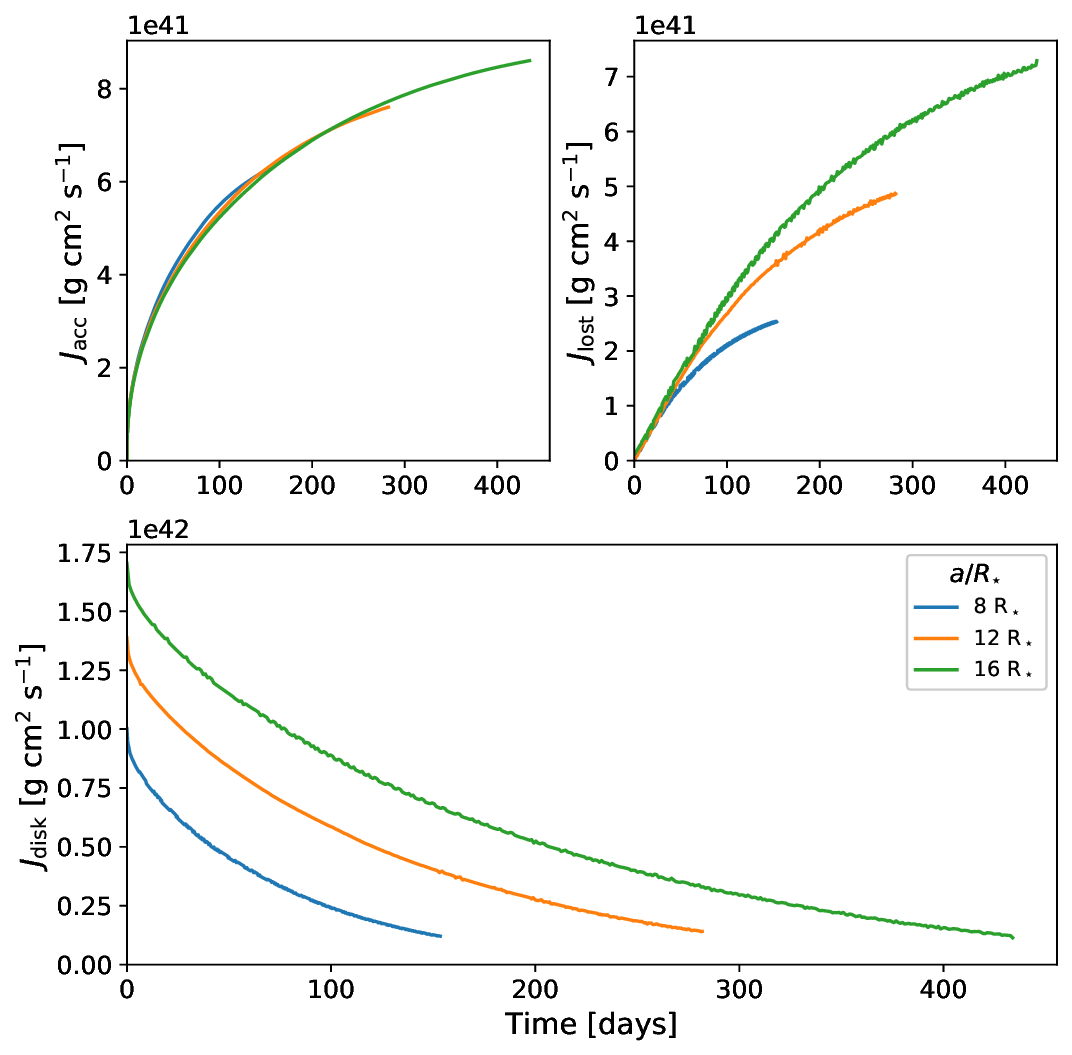}
\caption{Angular momentum evolution of three simulations with solar mass companions and orbital separations of 8, 12, and 16 $R_\star$. The total angular momentum in the disk, cumulative angular momentum re-accreted by the primary star, and cumulative angular momentum lost through the outer disk edge are shown in the bottom, top left, and top right panels, respectively. The evolution of each system is shown from $t=0$ until 20 $P_\mathrm{orb}$.}\label{fig:JdotEvol}
\end{figure*}

The time evolution of the angular momentum for three sample simulations is shown in Figure~\ref{fig:JdotEvol}. Each simulation has a solar mass companion and has orbital separations of 8, 12, and 16 $R_\star$, respectively, which are each evolved for 20 orbital periods. The total angular momentum within the disk, shown in the bottom panel, demonstrates that its loss follows an approximately exponential decay, with each disk exhibiting a distinct timescale.

Despite the substantial differences in initial angular momentum, all disks exhibited remarkably similar re-accretion rates throughout their evolution. As shown in the top left panel of Figure~\ref{fig:JdotEvol},the angular momentum accreted by the primary during the first 50 days varies by less than 1\% across the sample simulations, with only shows minor differences after this point. This consistency is expected, as the material near the stellar surface is largely unaffected by the gravitational influence of the secondary. Given that all disks share the same base density of 5~g/cm$^3$, accretion of the inner disk should therefore proceed at similar rates.

\begin{figure*}[t]
\centering
\includegraphics[width=0.8\textwidth]{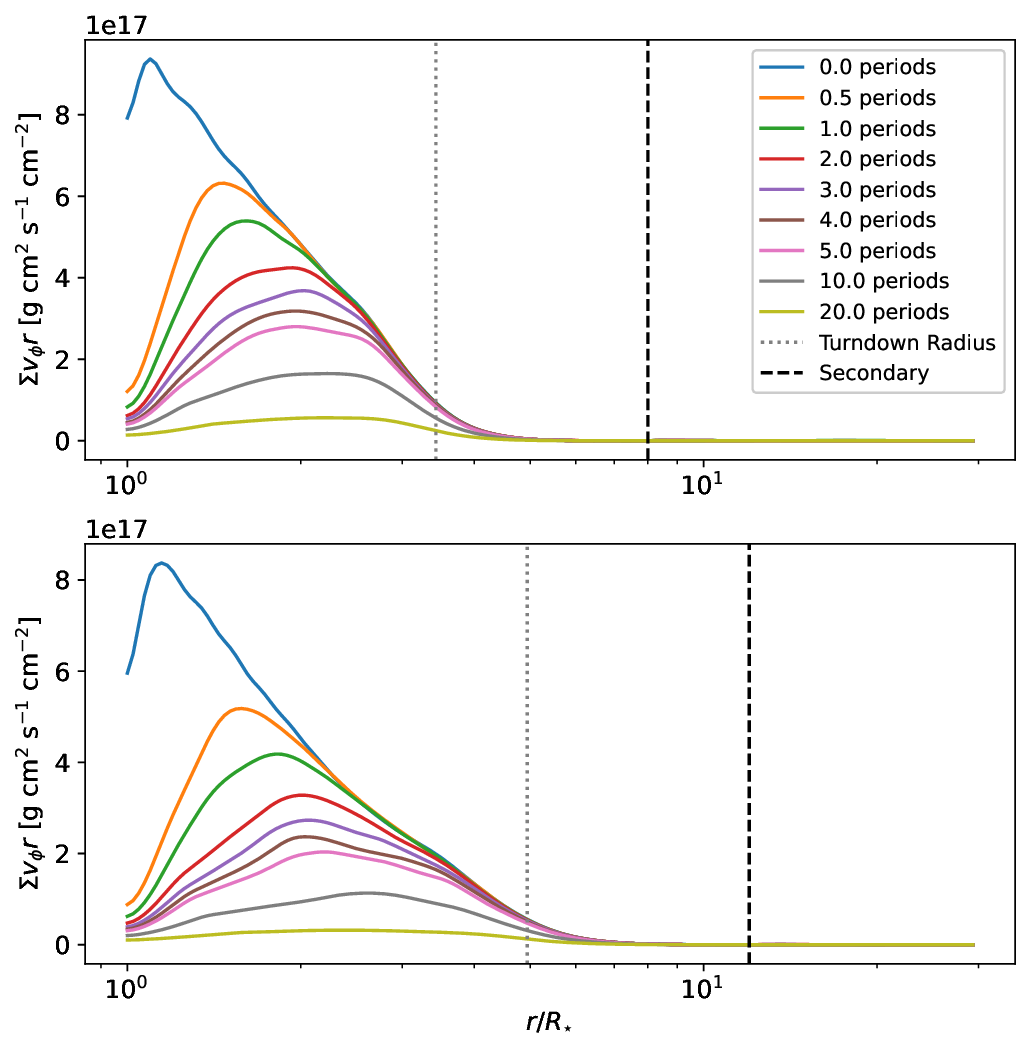}
\caption{Momentum density evolution of two different simulations with solar mass companions and orbital separations of 8 (top) and 12 (bottom) $R_\star$. The peak values are larger and closer to the stellar surface for the smaller disk, despite the total angular momentum in the disk being lower.}\label{fig:Jdens}
\end{figure*}

After approximately 50 days, however, the re-accretion rates of the smaller disks briefly surpass those of the larger disks. This phase is short-lived, with only a few thousand particles remaining in the disk afterwards, causing accretion to slow significantly. The temporary increase in smaller disks arises because their momentum density, $\Sigma v_\phi r$, peaks closer to the stellar surface and in greater magnitude than in larger disks. This structure is illustrated in  Figure~\ref{fig:Jdens}, which compares simulations with orbital separations of 8 and 12 $R_\star$, each with a solar mass companion. The inward-shifted peaks are a consequence of the shallower surface density profile induced in a close or massive binary through the mass accumulation effect. As the disks evolve, angular momentum is transported outward, causing material closer to the star to accrete first. Consequently, smaller disks whose peak momentum density lies nearer to the star reach and deplete their peak accretion region earlier.

In contrast, angular momentum loss from the outer disk edge varies significantly for each simulation, with larger disks losing angular momentum at higher rates. This is likely due to the lower gravitational influence from the primary at large radii, allowing the secondary to torque the material out of the disk with greater ease. Compounding this effect is that particles at large radii have proportionally larger specific angular momentum, and are therefore able to carry away more of it when flowing through the outer edge. 

Since the re-accretion rate is consistent across systems but the loss through the outer edge is not, this causes a higher percentage of angular momentum to be re-accreted onto the star for smaller disks. This could act as a mechanism to prevent close binaries from shedding their angular momentum while intermediate and wide binaries would lose angular momentum to the secondary and circumbinary disks more efficiently in a given dissipation cycle. In fact, this would remain the case even if the disks did not completely dissipate during the quiescent phase, as the diverging outflow rates would still cause more angular momentum to be lost for larger disks.

\begin{figure}[t]
\centering
\includegraphics[width=\columnwidth]{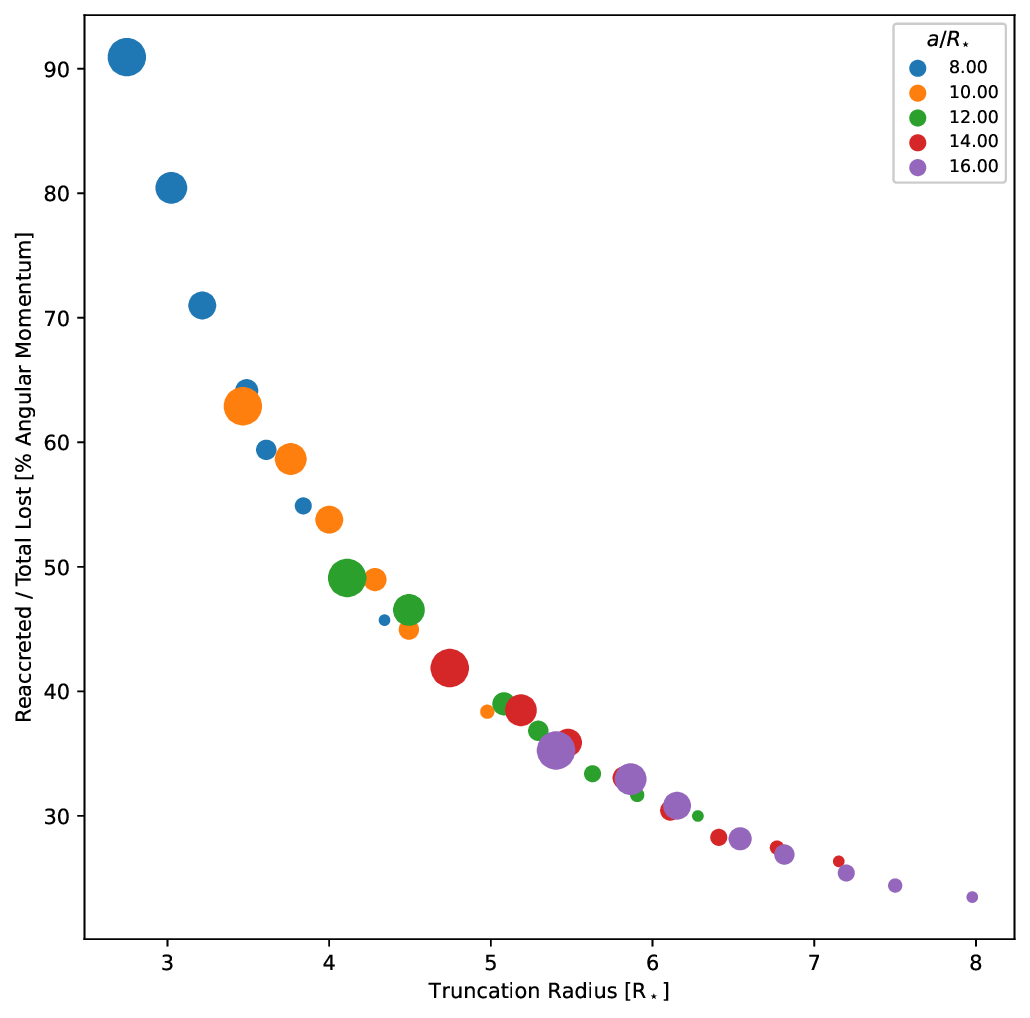}
\caption{Percentage of total angular momentum lost by the disk that is reaccreted by the Be star, plotted against the $v_\phi$ turnoff radius. Point color indicates orbital separation, and point area is proportional to the mass of the secondary star. This fraction has been measured after 20 $P_\mathrm{orb}$ in each simulation.}\label{fig:Jpaths}
\end{figure}

Figure~\ref{fig:Jpaths} shows the percentage of angular momentum lost from the disk due to reaccretion, plotted against the $v_\phi$ turnoff. The relationship follows an exponential decay, reaffirming that larger disks lose a greater fraction of angular momentum through outflow compared to smaller disks \textemdash not just those in systems with larger orbital separations. Since an empirical scaling law for the $v_\phi$ turnoff was established in Section~\ref{sec:ScaleLaw}, this curve serves as a numerical predictor for the efficiency of angular momentum loss based solely on the orbital characteristics. This relationship is particularly valuable because orbital characteristics, such as separation and mass ratio, are generally easier to measure observationally than the detailed structure of the disk itself. As a result, the $v_\phi$ turnoff scaling provides a practical tool for estimating angular momentum loss efficiency during dissipation without requiring direct tracking of disk material.

\section{Observables}
\label{sec:Observables}

\subsection{HDUST \& Simulation Details}

In order to predict the observational signatures of the dynamical models in Sections~\ref{sec:TruncRad} and \ref{sec:AngMom}, a three-dimensional radiative transfer code called \texttt{HDUST} is employed to generate spectroscopy and polarimetry data from selected SPH simulations at specific times. Below is a general description of \texttt{HDUST}, but the full details of the code are provided in the seminal paper \citet{car06}.

\texttt{HDUST} utilizes a Monte Carlo process to calculate the thermal structure of the disk through repeated propagation of photon packets through the circumstellar environment. An iterative scheme is employed to self-consistently solve and refine the ionization fraction and atomic level populations under the assumption of non-local thermodynamic equilibrium for the disk at a specified input time. These quantities are then used to predict polarimetric, photometric, and spectroscopic observables for a number of requested observer orientations.

In contrast to SPH, \texttt{HDUST} operates on a discrete three-dimensional grid, necessitating the transformation of physical quantities between representations. The densities and velocities calculated in SPH were converted to a meshgrid using an interface first presented in \cite{suf24}. We chose a spherical grid with 50 radial bins, 30 azimuthal bins, and 30 altitude bins, each uniformly spaced. Since \texttt{HDUST} does not currently include support for a secondary star, we chose to truncate the meshgrid at the secondary's Roche radius.

Accurately modeling observables from rapidly rotating stars requires accounting for rotational deformation and gravity darkening, as these effects significantly alter the stellar surface geometry and temperature distribution, thereby influencing the emergent radiation field \citep{von24}. Assuming the oft-used von Zeipel approximation for gravity darkening along with rigid rotation, the outward centrifugal force at the equator causes the star to become oblate according to,

\begin{equation}
    W = \sqrt{2\left(\frac{R_{eq}}{R_{pl}} - 1\right)},
\end{equation}

\noindent where $W$ is the ratio of the stellar rotation rate to the critical rotation rate, $R_{eq}$ is the equatorial radius, and $R_{pl}$ is the polar radius. \cite{riv13} shows $W=0.7$ to be the average for Be stars, which we will adopt here for our models.

Given the computational demands of \texttt{HDUST}, we restrict our analysis to a carefully selected subset of simulations for which observables are computed. To ensure adequate coverage of the range of disk sizes explored in this study, we focus on simulations with orbital separations of $a = 8,\, 12,\, 16 \, R_\star$, all with a companion mass of $M_s = 1.0M_\odot$. The outer boundary of the computational meshgrid is set at the Roche lobe of the secondary, corresponding to approximately $0.75\,a$. Observables are generated at time intervals of 0.0, 0.5, 1.0, 2.0, 3.0, 4.0, 5.0, and 10.0 orbital periods. This temporal sampling provides sufficient resolution to capture the rapid evolution during the early stages of disk dissipation, while also enabling analysis of the asymptotic behavior as the system approaches a more sparse state.

To investigate the full range of possible inclination angles, we select values of $i = 5^\circ$ (nearly face-on), $45^\circ$, and $85^\circ$ (nearly edge-on). In addition, we compute the observables for $i=70^\circ$, which has shown to produce the largest intrinsic polarization level \citep{woo96, hal13}. We also vary the azimuthal viewing angle, $\phi$, so that the disk is observed along the primary-secondary axis, $\phi=0^\circ$, or perpendicular to that axis, $\phi=90^\circ$.

\subsection{H$\alpha$ Analysis}

\begin{figure*}[t]
\centering
\includegraphics[width=0.8\textwidth]{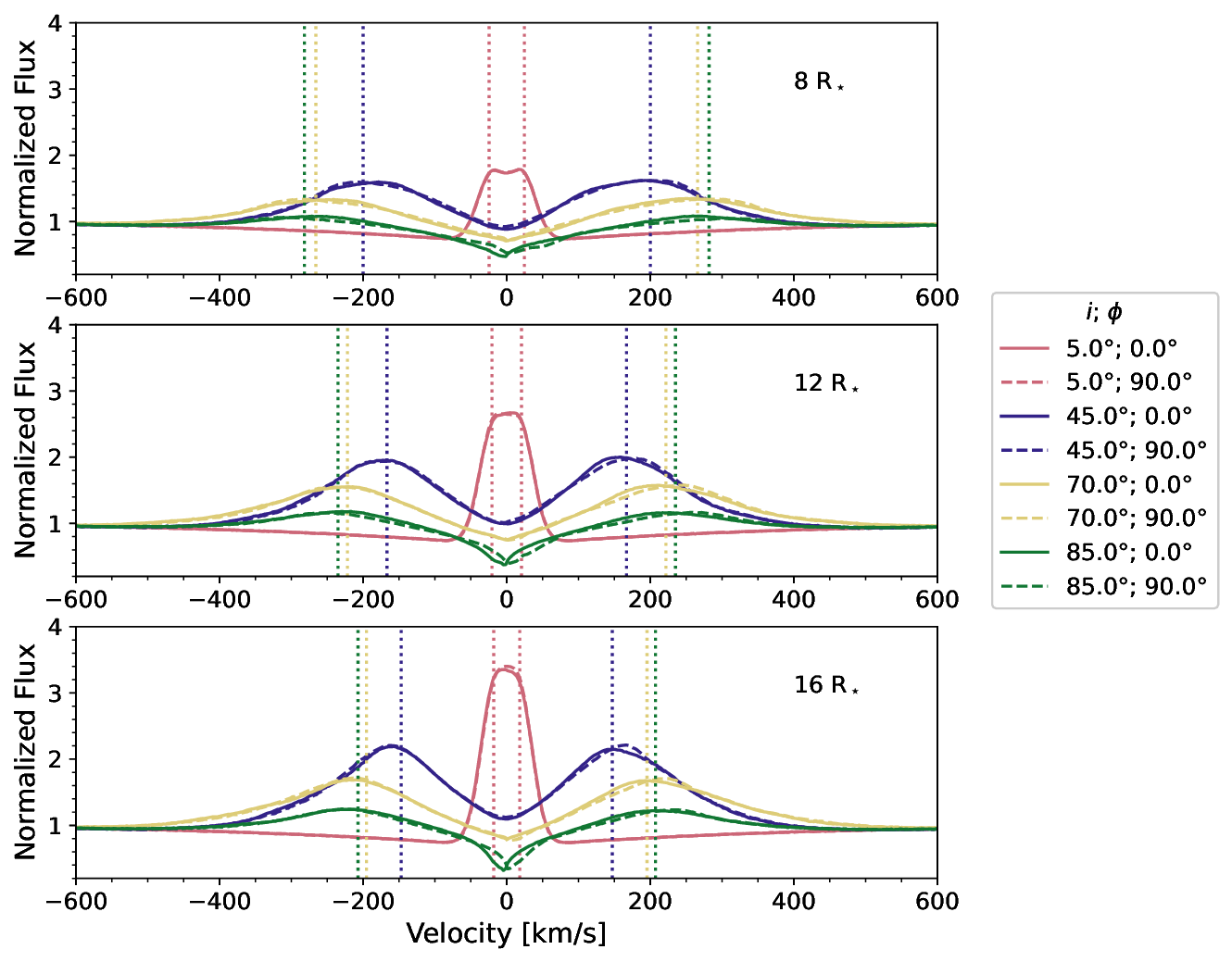}
\caption{Normalized H$\alpha$ line profiles for all \texttt{HDUST} simulations at $t=0$. Different viewing inclinations are represented by different colours, and different viewing azimuthal orientations are represented by either a solid line or dashed line for $0^\circ$ and $90^\circ$, respectively. Vertical dotted lines indicate the velocity at the $v_\phi$ turnoff radius for each simulation, corrected by $\sin i$.}\label{fig:Halpha}
\end{figure*}

The H$\alpha$ line shows the expected single-peak emission line for the low inclination case (near face-on) and double-peak for all other lines (near edge-on). The lines for each simulation at $t=0$ can be seen in Figure~\ref{fig:Halpha}, with the $\phi = 0^\circ$ simulations as the solid lines and the $\phi = 90^\circ$ as the dashed lines. In all models, changing the azimuthal viewing angle only changes the lines a very small amount. This reaffirms our implicit assertion that our models are close to rotationally symmetric, with the differences being the result of only minor density perturbations induced by the secondary. 

\begin{figure}[t]
\centering
\includegraphics[width=\columnwidth]{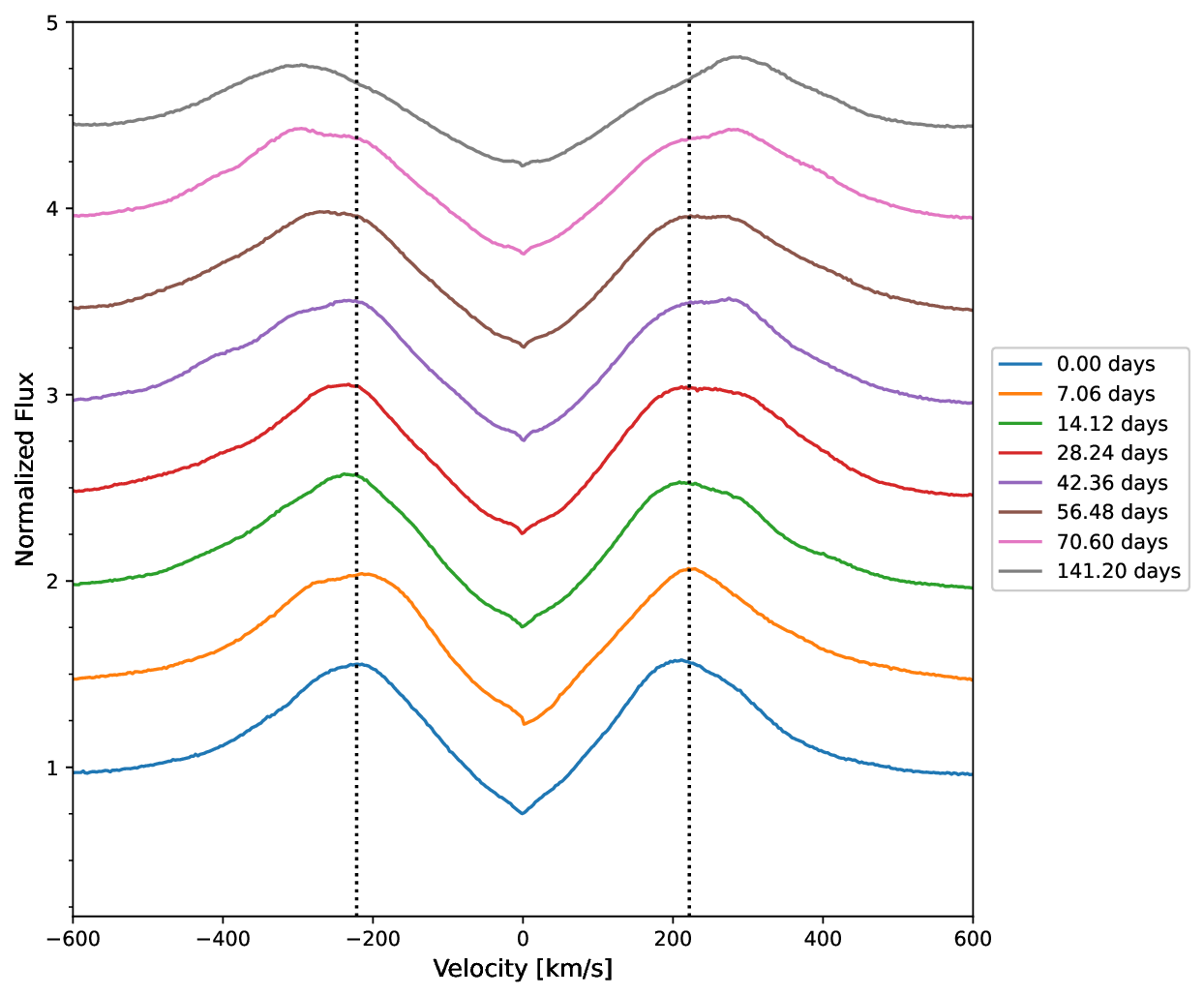}
\caption{Time-evolution of the normalized H$\alpha$ line profiles from the \texttt{HDUST} simulations of the $12R_\star$ system, viewed at an inclination of $i=0^\circ$ and azimuthal angle of $\phi=0^\circ$. Profiles are stacked vertically with time progressing from bottom to top, each offset by 0.5 units for clarity. Vertical dashed lines indicate the velocity at the $v_\phi$ turnoff, corrected by $\sin i$.}\label{fig:Halpha_diss}
\end{figure}

In Figure~\ref{fig:Halpha}, the vertical lines correspond to the azimuthal velocity at the $v_\phi$ turnoff, corrected by the line-of-sight component, $\sin i$, for their respective color. \cite{hua72} suggested that, where there is a double-peak structure, the peak-to-peak separation reflects the disk velocity at the outer disk edge. Our results show that the $v_\phi$ turnoff criterion provides a good approximation of the peak separation, consistent with this interpretation. However, as the disk dissipates, the peaks shift outward and decrease in intensity. This effect is illustrated in Figure~\ref{fig:Halpha_diss}, which shows the 12 $R_\star$ simulation viewed at $i=70^\circ$ and $\phi=0^\circ$. The outward shift corresponds to a loss of disk material, causing the outer regions to become optically thin and contribute less to the H$\alpha$ line profile. As these regions cease to significantly influence the emission, the outermost optically thick radius moves inward, toward regions of higher Keplerian velocity, resulting in the observed outward migration of the spectral peaks.

To enable a meaningful measurement of the $v_\phi$ turnoff, any observed system must therefore satisfy three key conditions. The first requirement is that the disk must be fully developed and in a steady state. This ensures that the peak-to-peak separation remains constant, eliminating ambiguity due to the timing of the measurement. This condition can be simply satisfied by selecting a photometrically stable Be star that has remained so for a sufficiently long time. The second requirement is that the disk remains optically thick out to its outer edge, ensuring that the H$\alpha$-emitting region encompasses the $v_\phi$ turnoff radius. This criterion is more challenging to verify, as disk base density is typically inferred through post-processing and modeling \citep{car06, kle17}. Nevertheless, it is a crucial step as an underdense disk may appear to truncate at smaller radii than it actually does. Fortunately, the base density chosen for the models in this section lies at the lower end of the typical range for B5 stars. As a result, similarly sized binary systems are likely to meet this optical depth requirement. Finally, to accurately determine the true azimuthal velocity at the $v_\phi$ turnoff, it is also necessary to infer the system's viewing inclination, as the observed quantity $v\,\sin{i}$ depends on both the intrinsic velocity and the observer.

Despite the restrictions, the ability to approximate the $v_\phi$ turnoff from the H$\alpha$ line profile can allow for observers to test the results of this paper. If both the binary parameters and H$\alpha$ peak-to-peak separation are known, this provides a direct method of testing the scaling law from Equation~\ref{eq:scalelaw}. More consequentially, a survey of photometrically-stable, circular binaries would allow for confirmation of our conclusion from Section~\ref{sec:AngMom}: that close, massive companions inhibit the loss of angular momentum through viscous transport.

\subsection{Polarimetry Analysis}

Polarization provides a powerful diagnostic of the densest, innermost regions of the circumstellar disk. In the context of accretion studies, processes occurring near the stellar surface are critical for constraining the mechanisms of mass and angular momentum exchange between the star and its disk. Polarimetric observations, therefore, offer unique insight into these regions, where direct measurements are otherwise challenging.

\begin{figure*}[t]
\centering
\includegraphics[width=\textwidth]{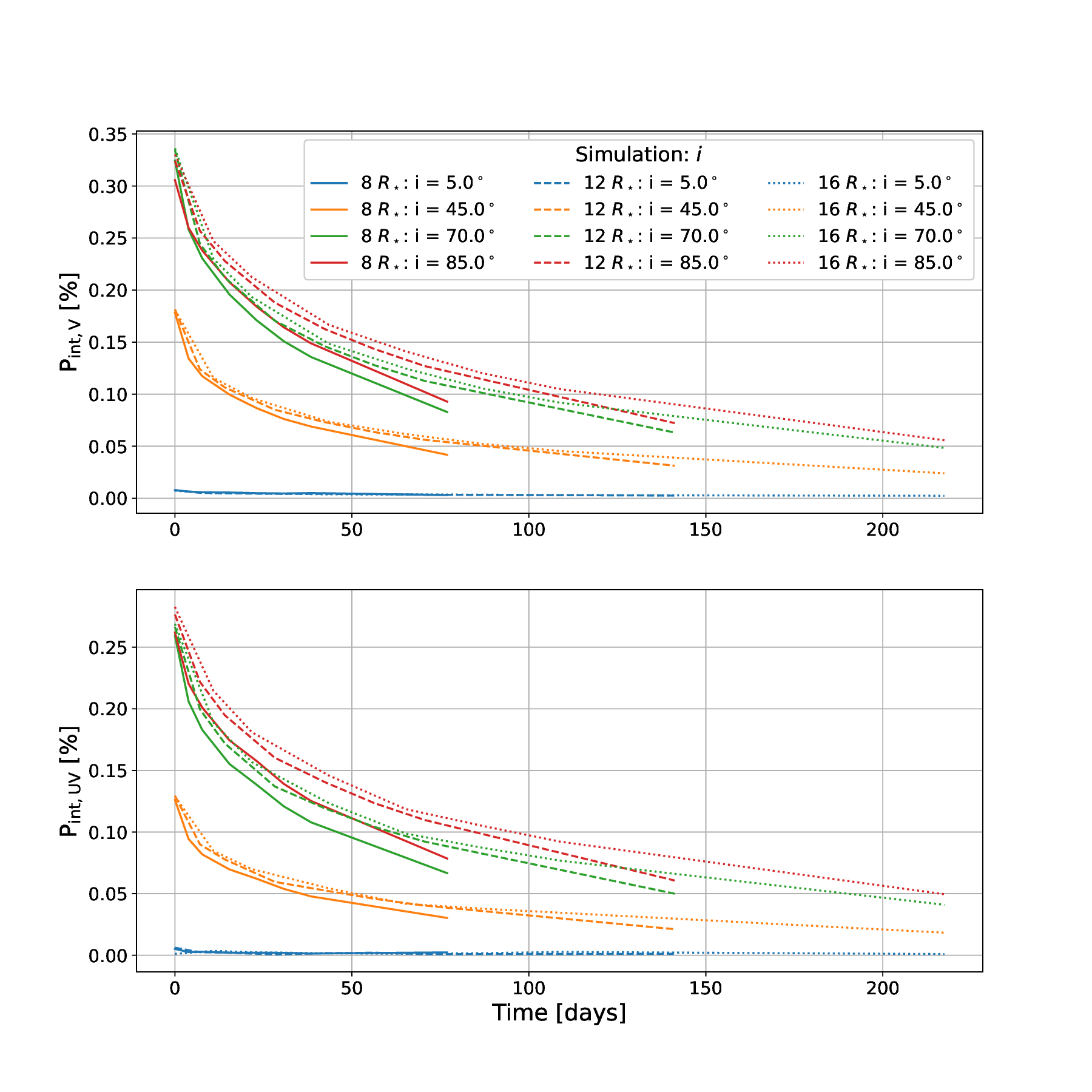}
\caption{Evolution of disk polarization. The top panel shows the V-band polarization, calculated as an average across the V-band (4500 - 6500 \AA). The bottom panel shows the UV polarization, calculated at an effective wavelength of 1500 \AA. Different simulations are distinguished by line style and viewing inclinations by color. All lines are for $\phi=0^\circ$.}\label{fig:Polarization}
\end{figure*}

Figure~\ref{fig:Polarization} shows the temporal evolution of the intrinsic polarization in the visible and UV regimes. The V-band polarization is computed as the average across the V-band wavelengths (4500 - 6500 \AA), while the UV polarization is evaluated at 1500 \AA, the fiducial wavelength of the Polstar mission. The \texttt{HDUST} models incorporate a comprehensive set of opacities including electron scattering, bound-free, free-free, and bound-bound processes. However, it is important to note that metal line blanketing is not included in these models, which likely causes the UV polarization to be overestimated, particularly in regions dense with spectral lines.

As one might expect, the polarization signatures at $t = 0$ are nearly identical across all disk models. This uniformity arises because polarization primarily probes the inner disk regions, which share the same base density in all cases. As such, unlike H$\alpha$ emission, polarization observation of a steady-state disk does not provide sufficient information to constrain the location of the $v_\phi$ turnoff in the system.

The beginning of dissipation coincides with a sharp drop in polarization within the first 20 days. Re-accretion of the dense inner disk occurs rapidly once it is no longer supplied with new material from the Be star, significantly reducing the number density of scatterers. Although this behaviour occurs in all simulations, the drop is less significant for larger disks which are able to better replenish the inner region with material from the outer disk. A similar explanation has been hypothesized to explain the sustained photometric behaviour of \cite{gho17} and \cite{rim18}, called the mass reservoir effect. 

The differences between simulations become more exaggerated as the disk continues to dissipate. After the initial drop, the polarization closely follows an exponential curve. Using another least-squares fitting algorithm, we can measure the mean timescale for the V-polarization to decay to half its initial value for the 8, 12, and 16 $R_\star$ simulations as 47, 62, and 83 days respectively, while the UV-polarization takes 46, 68, and 90 days respectively. Very little variation in this timescale is seen between each viewing inclination (i.e., less than a day), and so the average result has been taken.

One may expect that the polarization would imitate the angular momentum loss curve in the bottom panel of Figure~\ref{fig:JdotEvol} or the mass loss rate from the disk. However, measurement of these quantities show that the decay timescales are only similar for the smallest disk, and do not scale identically with $a$. Instead, the polarization decay timescales appear to vary proportional to $r_\mathrm{trunc}$. Since the $12R_\star$ and $16R_\star$ simulations have more material in the outer disk, and these regions supply the dense inner regions with material longer, which in turn sustains the polarization for longer compared to the $8R_\star$ simulation. Due to the broken power law profile of the surface density (see Figure~\ref{fig:SDconverge} for an example), the $v_\phi$ turnoff is a much more powerful indicator of a disk's ability to sustain itself than its orbital separation. A summary of the relevant time and length scales for each simulation is shown in Table~\ref{tab:decays}.

\begin{table*}
    \centering
    \begin{tabular}{||c|c|c|c|c|c||}
        \hline
        $a$ [$R_\star$] & $r_\mathrm{trunc}$ [$R_\star$] & $\dot{M}_\mathrm{disk}$ [days] & $\dot{J}_\mathrm{disk}$ & V-polarization [days] & UV-polarization [days] \\
        \hline \hline
        8 & 3.51 & 53 & 51 & 43, 44, 48 & 45, 45, 49 \\
        12 & 5.08 & 91 & 86 & 66, 67, 72 & 64, 67, 72 \\
        16 & 6.50 & 125 & 117 & 86, 86, 91 & 88, 89, 94 \\
        \hline
    \end{tabular}
    \caption{Characteristic length scales and timescales of each simulation. For V- and UV-polarization, the values are given for $i = 45^\circ,\, 70^\circ$, and $85^\circ$ respectively.}
    \label{tab:decays}
\end{table*}

This proportionality offers a promising method to independently estimate the $v_\phi$ turnoff using observable quantities. Since both V-band and UV polarization exhibit consistent relationships between the disk truncation radius and the polarization decay timescale, each wavelength regime can serve as an independent diagnostic. Cross-validating these predictions enhances confidence in the inferred disk dynamics and provides a robust test of the underlying physical models.

\section{Conclusions}

In this paper, we used the VDD model to study the dynamic and observable behaviour of dissipating Be disks in binary systems as angular momentum is removed from the circumprimary disk.

We introduced a novel analytical framework that uses the circular azimuthal velocity profile of Be star disks to define a physically motivated outer disk boundary. The ``$v_\phi$ turnoff" denotes the radius at which tidal torques begin to dominate over viscous circularization, causing the disk to deviate from a well-behaved circular velocity. By characterizing the disk in terms of this physically motivated boundary, the analysis of angular momentum transport is significantly streamlined: the azimuthal velocity becomes a fixed function of radius, and only the local density remains variable. Moreover, unlike criteria based on the surface density profile, the $v_\phi$ turnoff remains invariant during disk dissipation. Consequently, the limits of the angular momentum integral (Equation~\ref{eq:InnerDiskAngMom}) are rendered time-independent, providing a stable and consistent basis for modeling the disk’s dynamical evolution.

Forty three-dimensional smoothed-particle hydrodynamics simulations were run, each with a different combination of binary orbital radius and secondary star mass. The $v_\phi$ turnoff was measured in each system and the parameter space was fit using a least-squares algorithm to a template equation analogous to a Roche lobe approximation (Equation~\ref{eq:RL_alternative}). The analysis revealed that the $v_\phi$ turnoff scales sub-linearly with the semi-major axis, indicating that dynamical processes such as tidal torques and viscosity play a significant role in constraining the radial extent of the disk. Additionally, the $v_\phi$ turnoff exhibits a strong correlation with the Roche lobe radius of the primary star. This similarity suggests that the turnoff radius is governed by the topology of the gravitational potential, akin to the behavior of Roche lobe equipotentials. A key implication of this scaling relation is that the disk is consistently truncated at a radius smaller than the Roche lobe of the primary, reinforcing the role of tidal interactions in shaping disk structure.

After building to a steady-state distribution, each dynamical simulation was allowed to dissipate for 20 orbital periods, after which the simulations were terminated due to a low number of particles remaining in the disk. Particles which fell within the radius of the central star (re-accreted) or those transported beyond the disk's $v_\phi$ turnoff were tracked, and a thorough accounting of the angular momentum loss rate for both avenues was carried out. We found that the re-accretion rates for all systems are very similar, up until the disk becomes diffuse and the innermost region of the disk cannot be sufficiently resupplied with mass and angular momentum. In contrast, angular momentum loss through the outer boundary varies significantly between simulations, with the largest disks having the largest rates of angular momentum loss. As a result, the ratio of angular momentum lost due to re-accretion to the total angular momentum lost is well-correlated to the disk's $v_\phi$ turnoff.

Crucially, this finding suggests that disk dissipation cycles serve as an effective mechanism for angular momentum loss in Be stars primarily within intermediate to wide binary systems, or in those with low-mass companions. If one assumes that the mass and proximity of a companion do not fundamentally alter the star’s need to regulate its approach to critical rotation, then the active–quiescent duty cycle must adjust accordingly to compensate for variations in truncation efficiency. In systems where the disk is truncated at smaller radii, the star must either accumulate higher disk densities or undergo more frequent dissipation events to shed the same amount of angular momentum as systems with more extended disks. Alternatively, if no significant differences in behavior are observed, this would imply that two stars with comparable mass and rotation rates are nonetheless influenced at their surfaces by the presence of a companion. Such a result would place strong constraints on the formation and evolutionary pathways of Be stars in binaries.

To connect these theoretical findings with observables, a sample of the dynamical simulations were passed through the \texttt{HDUST} radiative transfer code, generating synthetic H$\alpha$ line profiles as well as diagnostics in the V-band and UV polarimetry. For a disk in steady state, it was found that the peak-to-peak separation of the H$\alpha$ line was well-approximated by the velocity at the $v_\phi$ turnoff of each disk. As a result, for a spectroscopically stable star with a known base density and viewing inclination, the truncation radius can be inferred observationally, independent of known binary parameters, and may therefore serve as a test of the proposed scaling law. A comparison of the polarization of each disk revealed markedly similar initial levels of polarization due to the identical base density at $t=0$. Further evolution of the disk causes the simulations to diverge, with smaller disks decaying faster than larger disks. The corresponding timescale was compared to several other time and length scales and was found to increase proportional to the $v_\phi$ turnoff, suggesting the possibility of another method to measure the disk size from observations.

In summary, this work details the inhibiting effect of a close binary companion on the angular momentum loss through the disk once in a quiescent phase. We accomplished this by identifying a new disk edge criterion, the $v_\phi$ turnoff, which is shown to be an advantageous tool in analyzing dissipating disks over other criteria such as surface density profiles, especially for analyzing angular momentum. Under this framework, it was shown that the rate angular momentum is torqued out of the disk is proportional to the size of the disk, and hence larger disks reaccrete a smaller fraction of their angular momentum during dissipation. This implies that closer binaries or those with larger companions are less able to efficiently remove angular momentum from their surface, which may impact the evolution or duty-cycle period and intensity of the Be star. Finally, these results are connected to a set of predicted observables, which are intended to allow for observational verification of our predictions.

Future research should broaden the range of parameters explored in this study and compare our results to observed Be populations. Of particular interest is the viscous coupling strength, which likely plays a critical role in the $v_\phi$ turnoff by serving as the primary mechanism for redistributing angular momentum from a tidally-torqued annulus. Conducting a similar analysis of angular momentum loss in misaligned or elliptical systems could also shed light on how Be stars either lose or retain their spin under varying conditions throughout their evolutionary stages. A systematic review of observed Be stars and their duty cycles can then be used in conjunction with the expanded framework to directly test the central claim of this paper: that close, massive companions inhibit the efficiency of angular momentum loss from the Be star. Possible candidates which are similar in size to our simulated systems are $\kappa$ Dra \citep{kle22}, $\varphi$ Per \citep{gie98}, and $\pi$ Aqr \citep{con25} which have all additionally been observed to enter periods of dissipation. Together, these avenues of investigation promise to deepen our understanding of angular momentum evolution in the interior of Be stars, and refine the theoretical foundations that underpin their complex dynamical behavior.

\section{Declarations}

CEJ received support from the NSERC Discovery Research Program. RGR received funding from the Natural Sciences and Engineering Research Council of Canada (NSERC) Postgraduate Scholarship - Doctoral. AuD has received research support from NASA through Chandra Award number TM4-25001A issued by the Chandra X-ray Observatory 27 Center, which is operated by the Smithsonian Astrophysical Observatory for and on behalf of NASA under contract NAS8-03060. AG acknowledges the support from the European Union (Project 101183150-OCEANS) and from Universidad Nacional de R\'io Negro, Argentina (Project 40-B-1186). JK was supported by grant GA \v{C}R 25-15910S. ACC acknowledges support from the Conselho Nacional de Desenvolvimento Cient\'ifico e Tecnol\'ogico (CNPq, grant 314545/2023-9) and the S\~ao Paulo Research Foundation (FAPESP, grants 2018/04055-8 and 2019/13354-1). 

Ethics declaration: not applicable.

\section{Author Contributions}
All authors contributed to the study conception and design. Peter Quigley conducted and analyzed all dynamical and radiative transfer simulations and contributed text to all sections. Carol E. Jones provided extensive feedback on the simulations and their analysis, and provided many relevant citations to supplement this work. Rina Rast analyzed the radiative transfer simulations and contributed expertise for the Observables section. All authors provided suggestions which shaped the analysis of this work. Alex C. Carciofi provided extensive support in analyzing the radiative transfer simulation data.  All authors commented on previous versions of the manuscript and have read and approved the final manuscript.

\bibliography{bibliography}

\end{document}